\documentclass[journal]{IEEEtran}

\ifCLASSINFOpdf
  \usepackage[pdftex]{graphicx}
\else
\fi

\usepackage{amsmath,amsfonts,amssymb}
\usepackage{times}
\usepackage{graphicx}
\usepackage{url}
\usepackage{booktabs}

\usepackage{array}

\usepackage{multicol}
\usepackage{multirow}

\newcommand{\tabincell}[2]{\begin{tabular}{@{}#1@{}}#2\end{tabular}}

\usepackage{url}

\begin{document}
%
\title{What, Where and How to Transfer in SAR Target Recognition Based on Deep CNNs}
%
%
%

\author{Zhongling~Huang,
        Zongxu~Pan,
        and~Bin~Lei
\thanks{This work was supported by the National Natural Science Foundation of China under Grant 61701478 and the Joint Training Program of University of Chinese Academy of Sciences.}
\thanks{Zhongling Huang is with School of Electronic, Electrical and Communication Engineering, University of Chinese Academy of Sciences, Huairou District, Beijing 101408, China. (e-mail: huangzhongling15@mails.ucas.ac.cn)}
\thanks{The authors are with the Key Laboratory of Technology in Geo-spatial Information Processing and Application System, Chinese Academy of Sciences, Beijing 100190, China, and also with Institute of Electronics, Chinese Academy of Sciences, Beijing 100190, China.}
}

\maketitle

\begin{abstract}
Deep convolutional neural networks (DCNNs) have attracted much attention in remote sensing recently. Compared with the large-scale annotated dataset in natural images, the lack of labeled data in remote sensing becomes an obstacle to train a deep network very well, especially in SAR image interpretation. Transfer learning provides an effective way to solve this problem by borrowing the knowledge from the source task to the target task. In optical remote sensing application, a prevalent mechanism is to fine-tune on an existing model pre-trained with a large-scale natural image dataset, such as ImageNet. However, this scheme does not achieve satisfactory performance for SAR application because of the prominent discrepancy between SAR and optical images. In this paper, we attempt to discuss three issues that are seldom studied before in detail: (1) what network and source tasks are better to transfer to SAR targets, (2) in which layer are transferred features more generic to SAR targets and (3) how to transfer effectively to SAR targets recognition. Based on the analysis, a transitive transfer method via multi-source data with domain adaptation is proposed in this paper to decrease the discrepancy between the source data and SAR targets. Several experiments are conducted on OpenSARShip. The results indicate that the universal conclusions about transfer learning in natural images cannot be completely applied to SAR targets, and the analysis of what and where to transfer in SAR target recognition is helpful to decide how to transfer more effectively. 
\end{abstract}


%
\IEEEpeerreviewmaketitle

\section{Introduction}
%
%
%
%
\IEEEPARstart{D}{eep} learning techniques, which automatically learn effective hierarchical features from the large-scale dataset, have been widely used in remote sensing data analysis in recent years. However, scarce labeled data, the biggest obstacle of applying deep learning to the field of remote sensing, still exists and significantly restricts the further development. Different from tasks that have millions of labeled samples in natural image fields, the training data in the remote sensing field is usually inadequate to train a deep network well. Instead of training a deep network from scratch with a few data, transfer learning, which aims to transfer knowledge from the source domain with a large-scale dataset to the target domain, provides an effective way to train a deep network with limited data. The most straightforward and commonly used trick is to fine-tune the network based on a pre-trained one.

Remote sensing data, mainly from optical (multi- and hyper-spectral) and synthetic aperture radar (SAR) sensors, are multi-modal with different imaging geometries and content. Penatti et al. \cite{17penatti2015deep} firstly indicated that the deep features can be generalized from everyday objects in daily images to objects in optical remote sensing images. Different kinds of convolution neural networks (CNNs), such as CaffeNet, AlexNet, VGG, trained on ImageNet, a natural image dataset, are tried to transfer to 3 bands optical remote sensing images classification, and achieve a remarkable performance \cite{21rs71114680}. Many subsequent literatures choose a variety of existing successfully pre-trained CNN models on ImageNet to transfer to various tasks, such as image registration \cite{5Wang2018A}, airplane detection \cite{13Chen2018End}, scene classification \cite{19marmanis2016deep,20zhao2017transfer}, image segmentation \cite{22fu2017classification} and image super-resolution \cite{yuan2017hyperspectral}, for both hyper-spectral and multi-channel remote sensing images. For optical remote sensing applications, transferring knowledge from natural images is prevalent since imaging mechanisms of both natural and optical remote sensing images are the same so that they can share some low- and mid-level features, such as those resemble either Gabor filters or color blobs. Apart from taking natural images as the source data, remote sensing data obtained from other platforms can also be used. Windrim et al. \cite{18windrim2018pretraining} transfers CNNs trained from certain hyper-spectral images (HSI) to classify other HSI from different satellite(aerial) platform. Similarly, Samat et al. \cite{4Samat2016Geodesic} transfers between training and validation data of hyper-spectral images with domain adaptation to weaken the statistical distribution difference. 

Due to different imaging mechanisms, approaches for the interpretation of optical remote sensing images cannot be directly used for interpreting SAR images in general. While transfer learning has begun to attract attention in optical remote sensing application recently, relevant study in SAR images has not caught up with yet. We just find a few studies in which transfer learning is applied to conquer the difficulty of lacking labeled SAR data to train a deep network. Yang et al. \cite{10Yang2016Moving} made the classifier learn the common knowledge among with different target-aspect angles of SAR targets via transfer learning. Malmgren-Hansen et al. \cite{25malmgren2017improving} proposed a generation approach on SAR data, and answered the question about how to transfer the knowledge from the simulated SAR data to the real one. Huang et al. \cite{huang2017transfer} indicated that features learned from a large amount of unlabeled SAR scene images via stacked convolutional auto-encoders are transferable to SAR target recognition task. To our best knowledge, there is not yet adequate evidence indicating whether the optical images can or cannot be transferred to SAR images with effect. Although several studies have attempted to explore how to transfer knowledge from optical to SAR images, even the rationality of this transfer is still under debate. Pros are as follows: Kang et al. \cite{8Kang2016SAR} utilized the intermediate layers of the pre-trained network on CIFAR-10 dataset as the feature extractor for classification of TerraSAR-X images and Wang et al. \cite{12Wang2017Combing} fine-tuned the VGG-16 model trained under natural images to detect ships in SAR images. While cons also exist, for example, Marmanis et al. \cite{24marmanis2017artificial} thought that initialization with the weights learned from optical images has little effect on classification of SAR data, simply because the distributions of optical images and SAR data are probably too different from each other to transfer even in low layers.

Considering the particularity of SAR images, especially the different imaging mechanisms between optical and SAR sensors, it’s not easy to transfer the features immediately from those successfully models which are often trained with natural image dataset \cite{16zhu2017deep}. The problem of transferring from other datasets to remote sensing data with large variations still remains to be solved and the transferability of trained networks to other imaging modalities needs to be further investigated \cite{1Zhang2016Deep}. In this paper, we will explore transfer learning focusing on SAR target recognition in a further way and try to discover more properties.

The contribution of this paper is to answer the following three questions about transfer learning via CNNs for SAR target recognition.

\subsubsection{What to Transfer} The network and source tasks should be both considered in transfer learning on SAR target recognition. A deeper well-trained network with a large-scale dataset generally has a stronger ability in extracting generic features, and the distance between source data and target data affects the transferability of features. We explore the influence of different source data and tasks, including optical images, SAR scene images, and SAR target dataset, as well as classification and reconstruction, and different architectures to show what network together with datasets should be transferred to SAR target recognition. Besides, we propose a transitive transfer method via multi-source data to improve the generality of features in layers significantly.

\subsubsection{Where to Transfer} The transferability of features varies from layer to layer in deep CNNs. Some are general meaning that they are applicable to other tasks, while the others are more specific to a particular task. Generally speaking, the transferability of features decreases from low-level to high-level. We analyze the generality and specificity of features in different layers with various source data when taking SAR target recognition as the target task, so as to decide which level of the features can be used as the off-the-shelf representation for the target task.

\subsubsection{How to Transfer} To transfer features that are specific to a particular task from the source task to the target task, we propose a method based on multi-kernel maximum mean discrepancy in domain adaptation, which combines the unsupervised and supervised learning to utilize the best of source and target data regardless of the labels. The approach increases the generality of features in task-specific layers, resulting in a stronger feature representation of the target data and a better performance in recognition.

\subsubsection{SAR Specific Model} We provide the SAR specific model pre-trained on a large-scale SAR land cover and land use dataset with a strong ability to extract spatial features of SAR images, which is validated to be well transferred to SAR targets, such as MSTAR and OpenSARShip datasets \cite{code}.

The rest of this paper is organized as follows. After a brief introduction of transfer learning and domain adaptation in Section \ref{2}, the proposed method is detailed in Section \ref{3}. Experiments and discussions are presented in Section \ref{4} to validate the effectiveness of the proposed method. Finally, the conclusions are drawn in Section \ref{5}.

\section{Related Work}
\label{2}
In this paper, we are interested in transfer learning on SAR target recognition. Consequently, we will introduce some typical literatures on SAR target recognition with transfer learning methods in this section firstly, and then followed by several related literatures about transfer learning and domain adaptation.

The simulated SAR images of vehicles with dense sampling of objects in different view angles are used for pre-training the CNN model to learn generic features that can be transferred to real SAR images in automatic target recognition (ATR) applications, proposed by Malmgren-Hansen et al. \cite{25malmgren2017improving} for the first time. However, the simulation of SAR images requires high technology but the technique is not so mature to simulate enough reliable models and difficult to popularize. On the other hand, Huang et al. \cite{huang2017transfer} found the features from unlabeled SAR scene images trained with a stacked convolutional auto-encoders are transferable to SAR targets. Although this is impressive and helpful under the case of lacking enough SAR targets but with adequate unlabeled SAR images, how generic or specific are the features from different source tasks transferred to SAR targets is still unknown. The transferability of features needs to be further explored and the discrepancy between source data and SAR targets should be fully taken into consideration to improve the performance of transfer learning.

Transfer learning, usually aiming at transferring knowledge from a large dataset known as the source domain to a small dataset called as the target domain \cite{27pan2010survey}, is widely popularized in deep convolutional neural networks based approaches. Yosinski et al. \cite{28yosinski2014transferable} discussed the transferability of features in deep neural networks, taking AlexNet trained on ImageNet as an example. They proposed a method to analyze how transferable the features are and found the generalization of features to other datasets and tasks apparently decreases as the layer goes deeper, leading to more specific features to a particular dataset or task especially in layer 6 and 7 which is widely applied in the subsequent studies \cite{29azizpour2015generic, DANlong2015learning, 7780678}. Although the co-adaptation of neurons between layers will bring out the optimization difficulty, fine-tuning the transferring features on the target dataset can disentangle this issue. Azizpour et al. \cite{29azizpour2015generic} investigated several influencing factors on transferability, including network structure, early stopping, fine-tuning, similarity between source and target tasks, etc. Among these factors, the similarity between source and target tasks is the most significant one to determine whether the learnt representation is generic or not. Considering the large domain discrepancy, Tan et al. \cite{32borgwardt2006integrating} proposed a transitive transfer learning method to transfer knowledge even when the source and target domains share few factors directly, with the aid of some annotated images as the intermediate domain to bridge them. Then they proposed a selective learning algorithm to transfer from face to airplane images which are totally different with each other \cite{33geng2011daml}. This also inspires us to think whether the intermediate task closer to SAR target recognition are capable to increase the feature generality. 

Domain adaptation approaches are often adopted to decrease the domain discrepancy between source and target tasks in transfer learning. Maximum mean discrepancy (MMD), a distance between embeddings of the probability distributions in a reproducing kernel Hilbert space, proposed by Borgwardt et al. \cite{32borgwardt2006integrating}, is used as the discrepancy metric between the source and the target in domain adaptation and transfer learning \cite{31pan2008transfer,33geng2011daml,34duan2009domain}. Long et al. proposed the deep adaptation network \cite{DANlong2015learning}, residual transfer networks \cite{RTNlong2016unsupervised} and joint adaptation network \cite{JANlong2017deep} successively based on the idea of domain adaptation using multi-kernel MMD metric to reduce the domain discrepancy between source and target, which inspired us to focus on learning transferable features in SAR target recognition. In most domain adaptation problems, the labeled source data and unlabeled target data are used, with common or similar categories but different distributions, such as Office-31 dataset which consists of 4,652 images within 31 categories collected from different environment variation of Amazon (downloaded from amazon.com), Webcam (taken by web camera) and DSLR (taken by digital SLR camera). In our case, however, the categories of SAR targets to be recognized are usually never seen before so the classifier should be retrained. Moreover, those methods of transferring among natural images are probably not applicable in SAR targets. In this paper, we will explore the specialized regulations and approaches specific to SAR target recognition.

\section{Methods}
\label{3}
According to the three questions of what, where and how to transfer in SAR target recognition that this paper prepare to explore, we will firstly elaborate the method of analyzing the transferability of features and then propose our approaches to make full use of the transferred features.

\subsection{Generic or Specific}
\label{3.1}
The features extracted from different layers of deep convolutional networks can be grouped into two categories, the generic feature and the specific one. Features with generality means they are capable to represent other dataset and those with specificity are closely related to the chosen data or tasks. In order to analyze the transferability of features in different layers on SAR target recognition case, we adopt the method of qualifying the generality versus specificity of features in each layer of a deep CNN \cite{28yosinski2014transferable}. Suppose there are $n$ different source tasks to transfer. Denote the $ith$, $jth$ source tasks as $S^i$, $S^j$, respectively, and the target task as $T$. For a network $N$ with $L$ layers, we would like to explore: 1) whether the features from the $kth$ layer are generic to the target task or specific to the source task. 2) From which layer does the transferability of features decline dramatically.

\begin{figure}[!t]
\centering
\includegraphics[width=8cm]{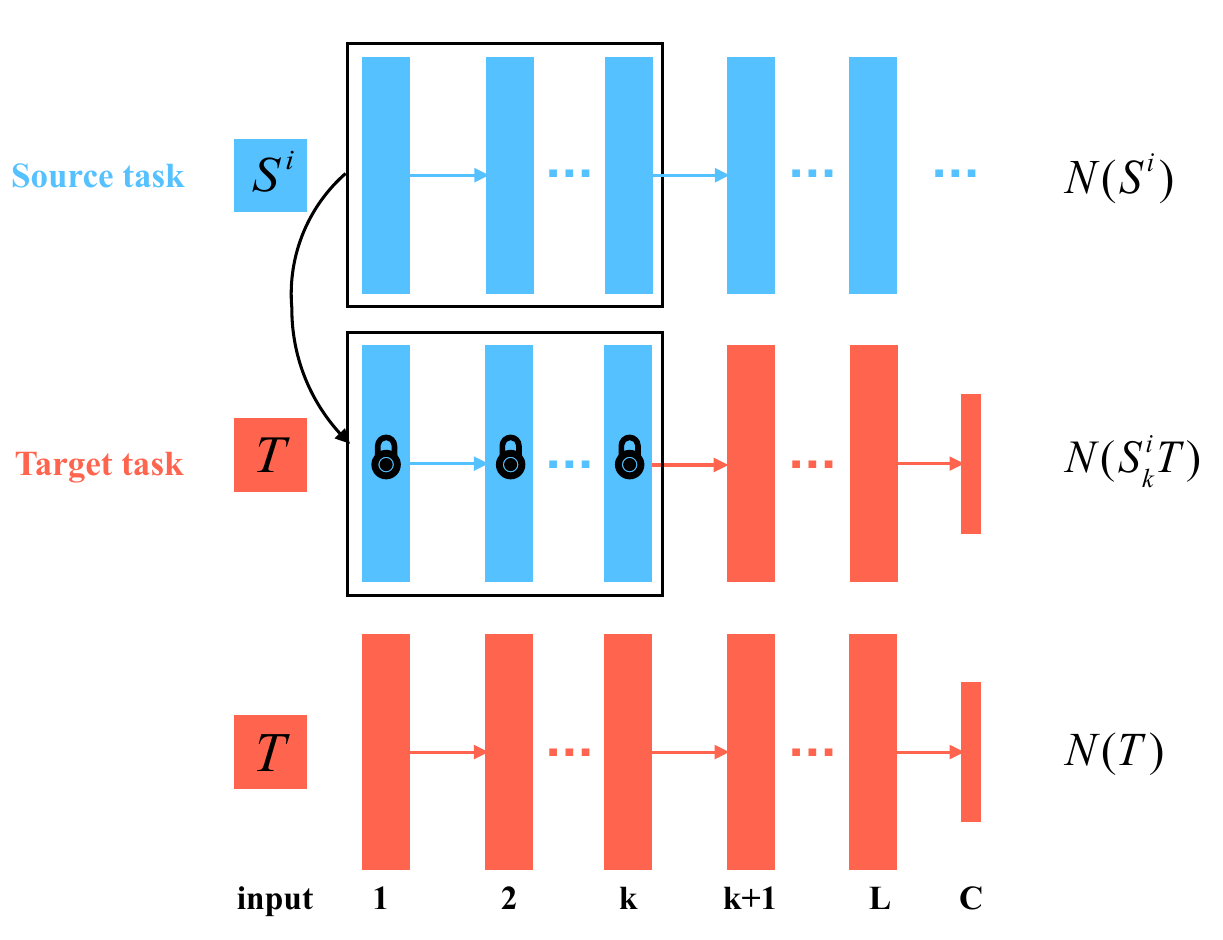}
\caption{The method of analyzing the generality of features. $N(S_{\mathrm{k}}^iT)$ is obtained by transferring and fixing the first $k$ layers of $N(S^i)$ then training the remaining randomly initialized layers on T. If the performance of $N(S_{\mathrm{k}}^iT)$ is better than the performance of $N(T)$, the features in layer k of $N(S^i)$ is considered as generic, specific otherwise.}
\label{Fig1}
\end{figure}

\begin{figure*}[!t]
\centering
\includegraphics[width=15cm]{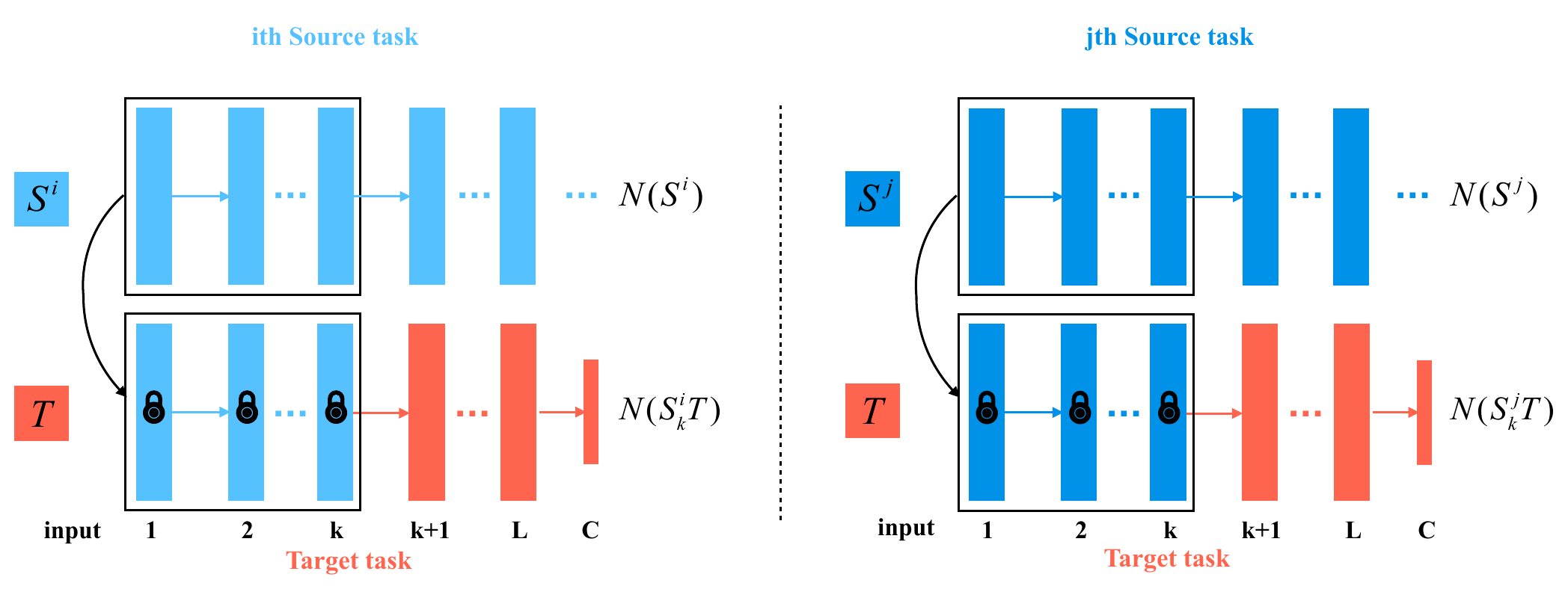}
\caption{To compare the transferability of different source tasks, different networks of $N(S_{\mathrm{k}}^iT)$ and $N(S_{\mathrm{k}}^jT)$ are obtained. If the performance of $N(S_{\mathrm{k}}^iT)$ is better than the performance of $N(S_{\mathrm{k}}^jT)$, $S^i$ appears more suitable to extract general features in layer k than $S^j$.}
\label{Fig2}	
\end{figure*}

Firstly, we train the network $N$ on source task $S^i$ from scratch, denoted as $N(S^i)$. Then the network is trained on $T$, with the $1 \sim k th$ layers copied from $N(S^i)$ and fixed as a feature extractor of the target task $T$ and the $k+1 \sim L th$ layers, as well as the classification layer $C$ randomly initialized, as shown in Fig. \ref{Fig1}. If the performance of this transferred network on $T$, denoted as $N(S^i_{\mathrm{k}}T)$ is better than the performance of the retrained network on $T$, denoted as $N(T)$, the features in layer $k$ are declared to be general. Otherwise, they are deemed to be specific to $S^i$. We compare the performance of $N(S^i_{\mathrm{k}}T)$ and $N(S^j_{\mathrm{k}}T)$ to evaluate the degree of generality of the $kth$ layer features from different source tasks $S^i$ and $S^j$, as shown in Fig. \ref{Fig2}. The results are given in Section \ref{4}.

\subsection{Transitive Transfer via Multi-Source}
\label{3.2}

\begin{figure}[!t]
\centering
\includegraphics[width=8cm]{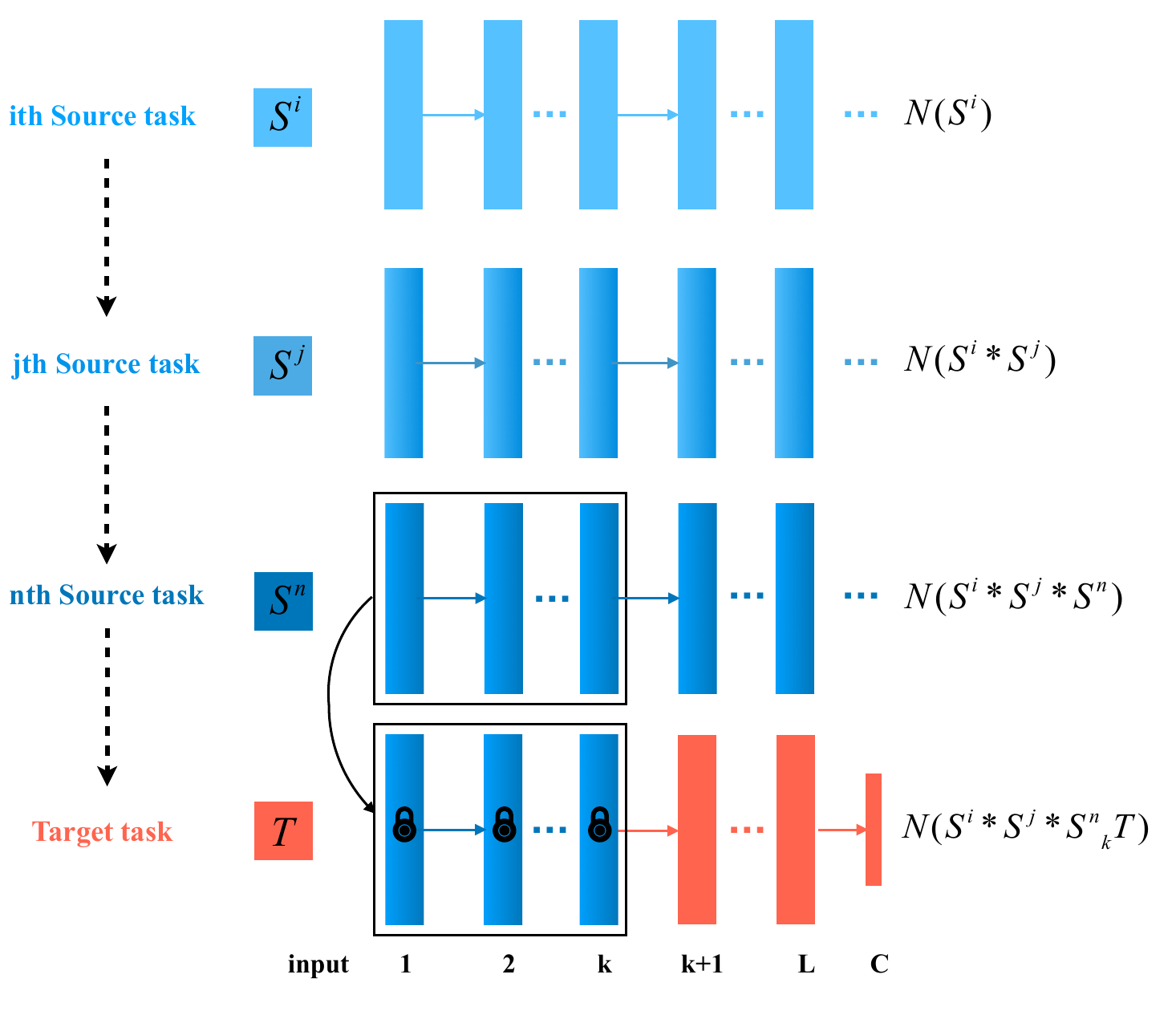}
\caption{Transitive transfer via multi-source from $S^i$ to $S^n$ and the first k layers transferred to $T$ at last.}
\label{Fig3}
\end{figure}

In this paper, we propose the transitive transfer via multi-source datasets. In the field of SAR image interpretation, various kind of tasks, such as image classification, reconstruction, target detection and recognition, are solved individually. Even for similar tasks, different problems usually do not cross paths with each other. Taking target recognition as an example, recognizing targets in optical images and SAR images, or recognizing different kinds of SAR targets such as airplanes and ships, are usually looked upon as different problems. Deep learning is a powerful tool to complete those tasks but training a new network for each task is time-consuming and data hungry for some tasks with limited labeled data. What if transitively transferring the knowledge task by task, especially from remotely similar task to similar one? Can it be helpful to enhance and enrich the ability of feature extraction on target dataset? In our method, as shown in Fig. \ref{Fig3}, given a network $N(S^i)$ trained on $S^i$, we simply fine-tune all layers on $S^j$ to fit the $jth$ source task, obtaining the network $N(S^i*S^j)$. Similarly, we get the $N(S^i*S^j…*S^n)$ with knowledge from source data $S^i$, $S^j$, … and $S^n$. And then we will analyze the transferability of features in each layer similar to Section \ref{3.1}. The results are given in Section \ref{4}.

\subsection{Transfer Learning with Domain Adaptation}
\label{3.3}
According to the previous analysis, specific features constrain the transferring among various tasks. To solve this issue, we propose a transitive transfer based method with domain adaptation to decrease the discrepancy between source and target task. Firstly, we will introduce the multi-kernel maximum mean discrepancy (MK-MMD) and then the two algorithms of the proposed method will be presented.

\subsubsection{Multi-Kernel Maximum Mean Discrepancy (MK-MMD)}
\label{3.3.1}
Maximum mean discrepancy (MMD) was firstly proposed by Borgwardt et al. \cite{32borgwardt2006integrating} as the discrepancy metric to compare the distributions based on two sets of data. In transfer learning, most domain adaptation methods are based on the MMD to narrow the gap between source and target domain. Suppose the distributions of the source data $\{x^s\}$ and target data $\{x^t\}$ are $p$ and $q$, respectively. For two dataset $\mathcal{D}_s=\{(x^s_i, y^s_i)\}_{i=1}^m$ and $\mathcal{D}_t=\{(x^t_j, y^t_j)\}_{j=1}^n$ with different distributions $p$ and $q$, their MMD is defined as

\begin{equation}
MMD(x^s,x^t)=\sup_{||\phi||_\mathcal{H}\leqslant1}\left[ E_{x^s \sim p}[\phi(x^s)] - E_{x^t \sim q}[\phi(x^t)] \right]_\mathcal{H}
\end{equation}
where $\phi(\cdot)$ denotes an element of a set of functions in the unit ball of a Reproducing Kernel Hilbert Space (RKHS) $\mathcal{H}$ and $E_{x^s \sim p}[\cdot]$ denotes the expectation of $x^s$ with the distribution $p$. In RKHS, the expectation is referred to as the embedding of $p$, and denoted as $\mu_p$ for short, that is

\begin{equation}
\mu_p=E_{x^s \sim p}[\phi(x^s)]
\end{equation}
As a result, the MMD can be regarded as a distance between embeddings of the probability distributions in a RKHS which represents a metric of source and target data. Furthermore, the square of MMD can be written as

\begin{equation}
\begin{split}
MMD^2(x^s,x^t) &= E_{x^s \sim p} \langle \phi(x^s),\phi(x'^s) \rangle _\mathcal{H} \\
& + E_{x^t \sim q} \langle \phi(x^t),\phi(x'^t) \rangle _\mathcal{H}  \\
& - 2E_{x^s \sim p, x^t \sim q} \langle \phi(x^s),\phi(x^t) \rangle _\mathcal{H}
\end{split}
\end{equation}
where the $\langle \cdot \rangle _\mathcal{H}$ denotes the inner product in RKHS $\mathcal{H}$ and the feature map $\phi(\cdot)$ can be associated with the kernel map $k(x^s,x^t)=\langle \phi(x^s),\phi(x^t) \rangle _\mathcal{H}$ in RKHS. Consequently, the empirical estimate of MMD can be given by

\begin{equation}
\begin{split}
MMD^2(\mathcal{D}_s, \mathcal{D}_t) & = \frac{1}{m^2} \sum_{i,j=1}^{m} k(x_i^s,x_j^s) + \frac{1}{n^2} \sum_{i,j=1}^{n} k(x_i^t,x_j^t) \\
& - \frac{2}{mn} \sum_{i,j=1}^{m,n} k(x_i^s,x_j^t)
\end{split}
\end{equation}
The kernel $k$ is usually defined as the convex combination of $U$ basis kernels,

\begin{equation}
k(x^s,x^t)=\sum_{u=1}^{U} \beta_u k_u(x^s,x^t),s.t.\beta_u \geqslant 0, \sum_{u=1}^U \beta_u=1
\end{equation}
and in our method the Gaussian kernel function is selected as the basis kernel.

In order to use mini-batch stochastic gradient descent (SGD) more easily and less time-consumingly in CNN, Gretton et al. \cite{40gretton2012optimal} proposed the unbiased estimate of MK-MMD with linear complexity which gives an approximation of a summation form. Given a quad-tuple $z_i=(x_{2i-1}^s, x_{2i}^s, x_{2i-1}^t, x_{2i}^t)$, by supposing $m=n$, the square of MMD can be rewritten as

\begin{equation}
\label{equ1}
MMD^2(\mathcal{D}_s, \mathcal{D}_t)=\frac{2}{m}\sum_{i=1}^{m/2}h(z_i)
\end{equation}
where

\begin{equation}
\begin{split}
\label{equ2}
h(z_i)= & k(x_{2i-1}^s, x_{2i}^s) + k(x_{2i-1}^t, x_{2i}^t) \\
& - k(x_{2i-1}^s, x_{2i}^t) - k(x_{2i}^s, x_{2i-1}^t)
\end{split}
\end{equation}

\subsubsection{Deep Domain Adaptation Based on Transitive Transfer with Multi-Source}
\label{3.3.2}
In our method, we will choose a variety of source tasks with diverse similarity to the target task to assist recognizing some new types of SAR targets by transitive transfer learning from distant to similar. Given a set of source tasks and arrange them in ascending order according to the similarity with the target task ${S^1,S^2,…,S^n}$. We pre-train and fine-tune the network as proposed in Section \ref{3.2} and analyze the transferability of features in each layer to see where the generality drops fiercely. Suppose the first $k$ layers have the strong ability to extract general features of target data, denoted as the off-the-shelf layers, and the $k+1 \sim L$ layers are more specific than the previous layers, denoted as the adaptation layers. Since the $N(S^1*…*S^n)$ is fine-tuned on $S^n$ at last, we only adapt the datasets of $S^n$ and $T$.

In the popular domain adaptation methods \cite{DANlong2015learning,JANlong2017deep,35tzeng2014deep,36ganin2015unsupervised}, the source data and target data share the same set of categories but with different probability distributions with the target data all unlabeled. The classification loss of source data and MMD between source and target data are combined to back-propagate to decrease the discrepancy, and then the target data can be classified into categories directly. In our case, however, the types of SAR targets to be recognized are never seen before and the classification layer should be retrained. In this paper, we proposed two algorithms and will have an elaborate discussion in Section \ref{4} on how to choose appropriate algorithm in different scenarios. 

\begin{figure}[!t]
\centering
\includegraphics[width=8cm]{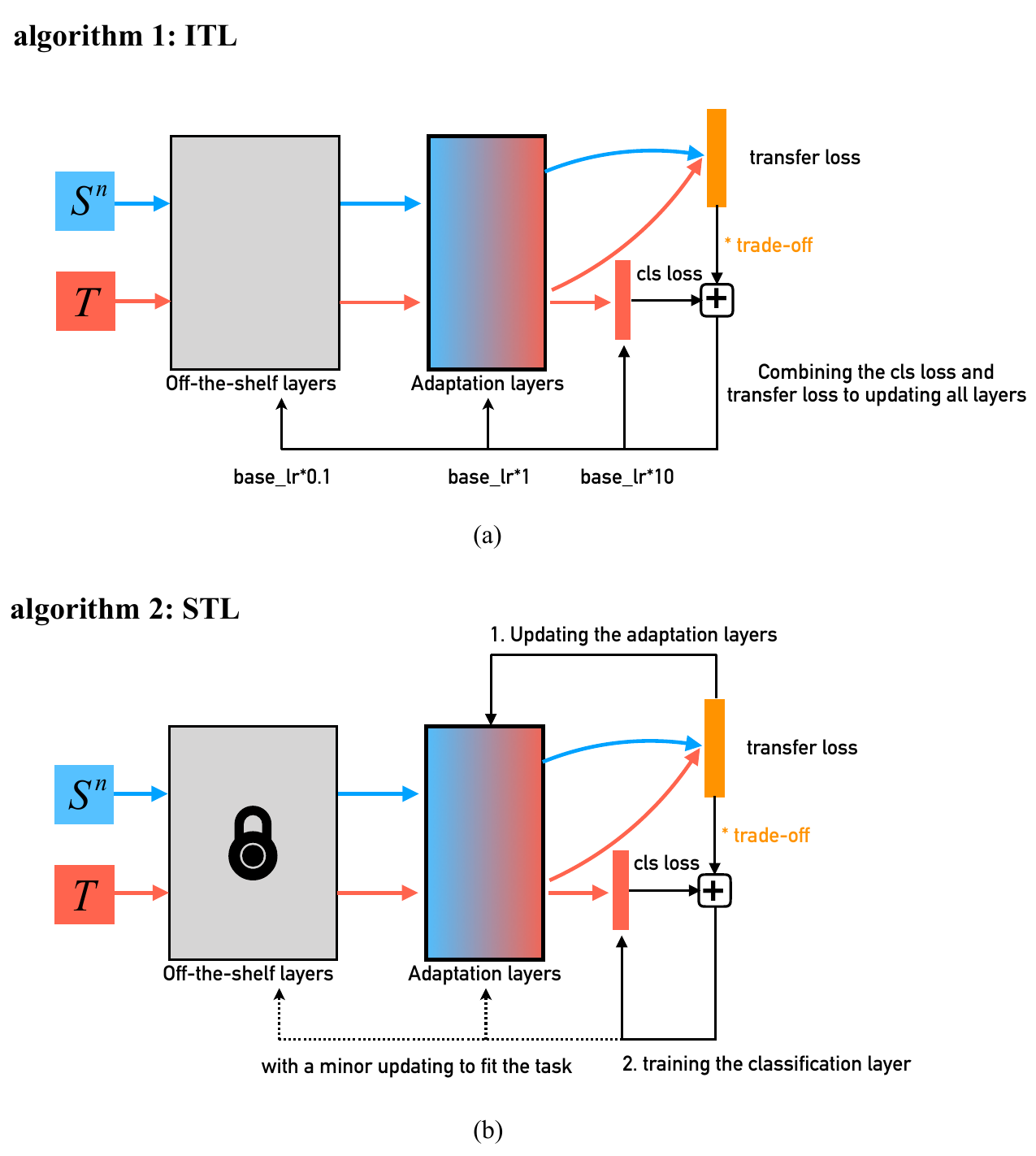}
\caption{The ITL and STL training algorithm based on domain adaptation. (a) presents the ITL algorithm which integrates the transfer loss in adaptation layers and the classification loss to back-propagate. (b) presents the STL algorithm to update the adaptation layers with transfer loss only and then train the classification layer and slightly fine-tune the off-the-shelf layers.}
\label{Fig4}
\end{figure}

Firstly, an integrated learning algorithm which combines the classification and domain adaptation is proposed as ITL, shown in Fig. \ref{Fig4}(a). Given a mini-batch of a quad-tuple of $\{ x_{2i-1}^s, x_{2i}^s, x_{2i-1}^t, x_{2i}^t \}$ as the input to the network, the transfer loss in the $lth$ adaptation layer is calculated by Equation. \eqref{equ1} and \eqref{equ2}, denoted as 

\begin{equation}
\label{mmd_l}
mmd_{l}(x_{2i-1}^s, x_{2i}^s, x_{2i-1}^t, x_{2i}^t)
\end{equation}
where $l=k+1,...,L$. The classification loss of target data is calculated by the standard Softmax loss, denoted as $\mathcal{L}_C(x_{2i-1}^t,y_{2i-1}^t; x_{2i}^t,y_{2i}^t;\theta_C)$ where $\theta_C$ represents the category classifier. The network is trained by minimize the total loss of

\begin{equation}
\begin{split}
& \mathcal{L}_C(x_{2i-1}^t,y_{2i-1}^t; x_{2i}^t,y_{2i}^t;\theta_C) \\
& + \lambda \sum_{l=k+1}^L \alpha_l mmd_{l}(x_{2i-1}^s, x_{2i}^s, x_{2i-1}^t, x_{2i}^t)
\end{split}
\end{equation}
where $\lambda$ denotes the trade-off between transfer loss and classification loss and $\alpha_l$ denotes the weight of transfer loss in each adaptation layer. In ITL algorithm, the transfer loss in adaptation layers are only added as a regularizer to classification and $\lambda$ is a dynamic parameter in the training process to keep a good balance on transfer loss and classification loss, especially at the later stage in training, $\lambda$ should be reduced by 0.1 to get a better trade-off. The setting of $\alpha$ depends on the transferability of each adaptation layer. Generally, the learning rate of the off-the-shelf layers should be smaller and the classification layer larger than the adaptation layers.

Secondly, considering the transfer loss and the classification loss are mutually interactive and restrictive when combined to optimize the parameters of the network, we propose a two-step training algorithm, namely STL as shown in Fig. \ref{Fig4}(b). In the first step of training the adaptation layers, the off-the-shelf layers are frozen because of the generality of representing the target data which also lowers the computational cost of optimizing the parameters. The transfer loss calculated by Equation. \eqref{mmd_l} is used to train the adaptation layers, aiming at decreasing the feature discrepancy in specific layers. Then the classification loss combined with the transfer loss is minimized to train the classification layer, with a minor updating in the off-the-shelf and adaptation layers. In the second step, $\lambda$ is reduced by 0.1 than the first step to make the transfer loss play a subordinate role as a constraint term.

\section{Experimental Results and Discussion}
\label{4}

\subsection{Datasets and Tasks Description}
\label{4.1}

In our experiments, we analyze the transferability of features using different source tasks and networks, and evaluate the proposed method on the target task, OpenSARShip recognition. The alternative source datasets / tasks contain ImageNet, TerraSAR-X images and SAR targets of MSTAR, for classification or reconstruction. Here are the brief descriptions of these datasets and tasks.

\subsubsection{ImageNet for Classification}
ImageNet is a well-known large-scale dataset of natural images in computer vision, providing the most comprehensive and diverse coverage of the image world \cite{37deng2009imagenet}. It contains 3.2 million labeled images over 5247 categories, over 600 images for each category on average. Generally, a subset of the large hand-labeled ImageNet dataset with 1.2 million images in 1000 object classes is considered as the benchmark to train the deep networks in the ImageNet Large Scale Visual Recognition Challenge (ILSVRC) \cite{38Russakovsky:2015:ILS:2846547.2846559} where the remarkable deep CNNs for image classification, such as AlexNet \cite{krizhevsky2012imagenet}, GoogLeNet \cite{szegedy2015going} and ResNet \cite{he2016deep}, are proposed in 2012, 2015 and 2016, respectively.

\subsubsection{TerraSAR-X Images for Classification and Reconstruction}
Firstly, we collect over 50,000 SAR image slices without annotation. These SAR slices are randomly cropped from SAR scene images covering various landscapes from TerraSAR-X, a German Earth-observation satellite which provides high-quality and precise earth observation data of 3 m resolution with StripMap mode. With rich texture information in those unlabeled SAR slices, a deep stacked convolutional auto-encoders is trained to reconstruct the slices, generating a series of hierarchical convolution layers capable of extracting efficient features.

\begin{figure}[h]
\centering
\includegraphics[width=9cm]{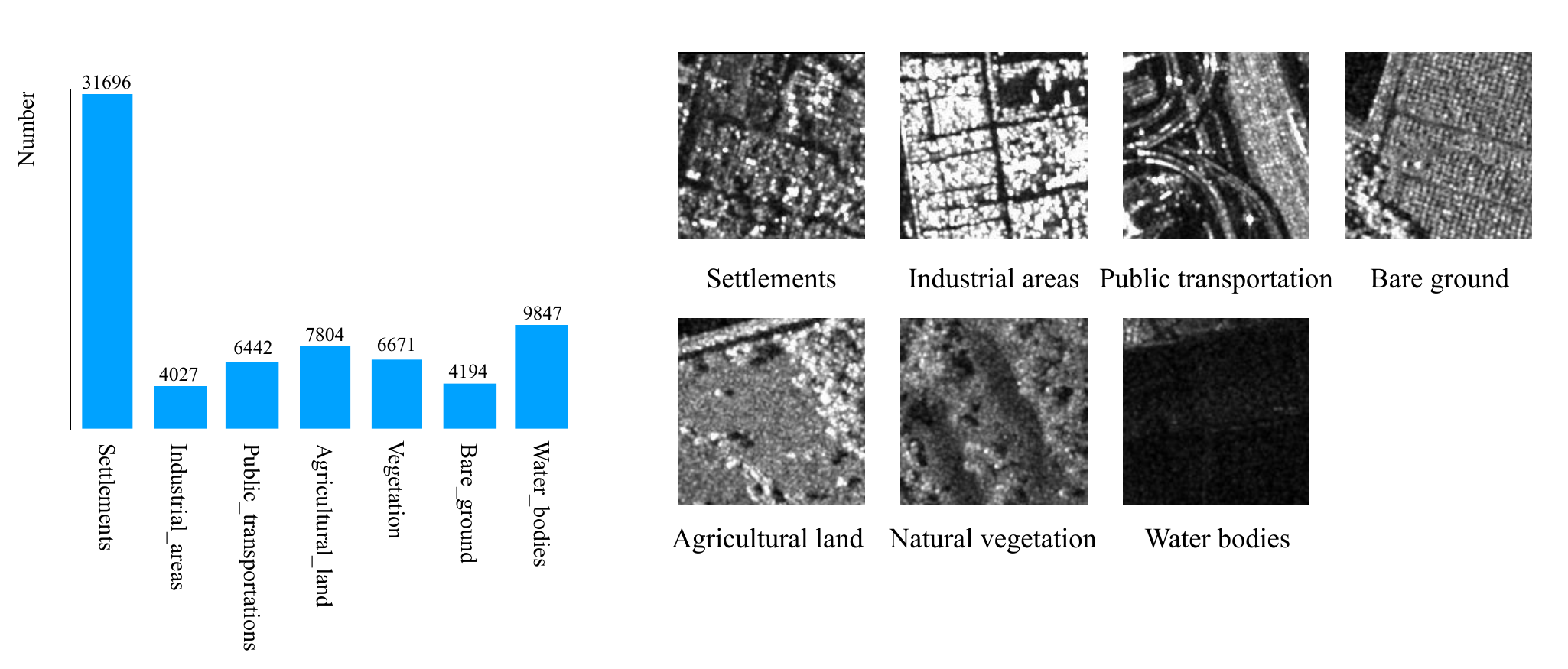}
\caption{The annotated TerraSAR-X land cover dataset with 7 categories out of 9 in level 1.}
\label{Fig6}
\end{figure}

Besides, a high-resolution SAR land cover annotated dataset \cite{dumitru2016land}, collected from TerraSAR-X horizontally polarized (HH), multi-look ground range detected (MGD) products, is applied for SAR land cover classification task in our experiments. The selected SAR images were taken in High Resolution Spotlight mode with the pixel spacing of 1.25 m, acquired with an incidence angle between 20$^{\circ}$ and 50$^{\circ}$, and with descending and ascending pass directions. Covering 288 full scenes of urban and non-urban areas all over the world, such as cities in Africa, Asia, Europe and some ocean areas, this dataset is hierarchically annotated of 3 levels, 150 categories and more than 100,000 patches. In our experiments, 7 categories with a high-level annotation of \textit{Settlements, Public transportation, Industrial areas, Agricultural land, Natural vegetation, Bare ground and Water bodies} are applied for classification, as shown in Fig. \ref{Fig6}.

\subsubsection{MSTAR for SAR Target Recognition}
The Moving and Stationary Target Acquisition and Recognition (MSTAR) public release dataset \cite{mstar} collected by Sandia National Laboratory SAR sensor platform contains 10 categories of military vehicles: the T72, BTR70, BMP2, 2S1, BRDM2, BTR60, D7, T62, ZIL131 and ZSU23, with the resolution of 1 ft on X-band. Those targets chips acquired at depression angle of 17$^{\circ}$ are usually used as the training data, 15$^{\circ}$ as the testing data to evaluate the SAR target recognition algorithms. Details of MSTAR dataset for 10-category SAR target recognition are shown in Table \ref{table2}.

\begin{table*}[!t]
	\caption{MSTAR Dataset}
	\label{table2}
	\centering
	\begin{tabular}{cccccccccccc}
		\toprule
		\textbf{Category}	& \textbf{2S1}	& \textbf{BMP2}	& \textbf{BRDM2}	& \textbf{BTR60} & \textbf{BTR70}	& \textbf{D7}	& \textbf{T62} & \textbf{T72} & \textbf{ZIL131} & \textbf{ZSU23}	& \textbf{Total}\\
		\midrule
		\textbf{17$^{\circ}$}		& 299	& 233	& 298	& 256	& 233	& 299	& 299	& 232	& 299	& 299	& \textbf{2747} \\
		\textbf{15$^{\circ}$}		& 274	& 195	& 274	& 295	& 196	& 274	& 273	& 196	& 274	& 274	& \textbf{2425} \\
		\bottomrule
	\end{tabular}
\end{table*}

\subsubsection{OpenSARShip for SAR Target Recognition}

\begin{figure}[!t]
\centering
\includegraphics[width=6cm]{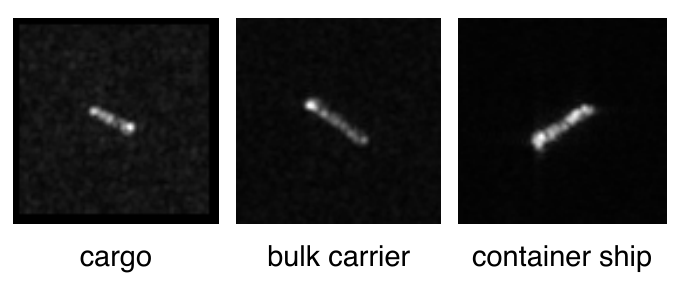}
\caption{Three elaborated types of OpenSARShip}
\label{Fig5}
\end{figure}

Huang et al. \cite{26huang2018opensarship} present a SAR ship dataset of Sentinel-1, containing 11346 ship chips from 41 Sentinel-1 SAR images. The dataset provides the Single Look Complex (SLC) and the Ground Range Detected (GRD) product of the IW mode, with polarization of VH and VV, as well as four formats of amplitude values, visualized data in gray scale, visualized data in pseudo-color and radiometric calibrated data. OpenSARShip contains 17 types of ships, such as Cargo, Tanker, Passenger, Tug, etc, but unbalanced numbers in each type (8470 in Cargo and 4 in Towing for example). There are 5 elaborated types in Cargo, naming Cargo, Container Ship, Bulk Carrier, General Cargo and Other Cargo. In order to evaluate the method with limited target data and balance the training numbers of each category, we select the elaborated types of Cargo, Container Ship and Bulk Carrier of GRD mode (with resolution of 10 m) and VV polarization in our experiments, filtering those ship chips with the size larger than 70 $\times$ 70 pixel to ensure the sufficient image information. The details are shown as Table \ref{OpenSARShip} and Fig. \ref{Fig5}.

\begin{table*}[!t]
\caption{OpenSARShip Dataset in Our Experiments}
\label{OpenSARShip}
\centering
\begin{tabular}{ccccc}
\toprule
\textbf{Elaborated Type} & \textbf{Cargo} & \textbf{Bulk Carrier} & \textbf{Container Ship} & \textbf{Total} \\
\midrule
\textbf{train} & 100 & 100 & 100 & \textbf{300} \\
\textbf{test} & 79 & 132 & 135 & \textbf{346} \\
\bottomrule
\end{tabular}
\end{table*}

\subsection{What to Transfer}
\label{4.2}
In this section, we will discuss how the different networks and source tasks affect the transferability of features in SAR target recognition and then apply the conclusion to our subsequent experiments. Azizpour et al. \cite{29azizpour2015generic} indicated that over-parameterizing the network by increasing the width (number of parameters at each layer) and depth (the number of convolution layers) can improve the performance on other datasets in transfer learning when they are close to the source tasks but may harm the transferability of features to distant target tasks. According to the previous researches, it's important to select the appropriate network and source task to transfer in SAR target recognition problem.

Aiming at recognizing the SAR ship targets of OpenSARShip dataset with only hundreds of labeled images, the first thought would be transferring layers from a close dataset such as MSTAR. We will discuss different networks pre-trained on MSTAR dataset in \ref{4.2.1}. Besides, we will explore how other source data or tasks, such as ImageNet classification, SAR images reconstruction, and SAR land cover classification, perform on transferring to SAR target recognition in \ref{4.2.2}, as well as the transitive transfer method using multi-source tasks.

\subsubsection{What Network}
\label{4.2.1}
 
 Three networks of A\_ConvNet \cite{chen2016target} which has the state-of-the-art performance on MSTAR recognition, H\_Net \cite{huang2017transfer} which is also well-performed on MSTAR using the stacked convolutional auto-encoders to learn hierarchical layers with unlabeled SAR images and transfer to SAR targets, and AlexNet \cite{krizhevsky2012imagenet} which is the breakthrough in large-scale image classification with deep neural network, are explored in this section. With more than 90\% of parameters in fully-connected layers, we only use the convolution layers of AlexNet due to the data scale in SAR targets, denoted as AlexNet\_Conv.

\begin{figure}[!t]
\centering
\includegraphics[width=6.5cm]{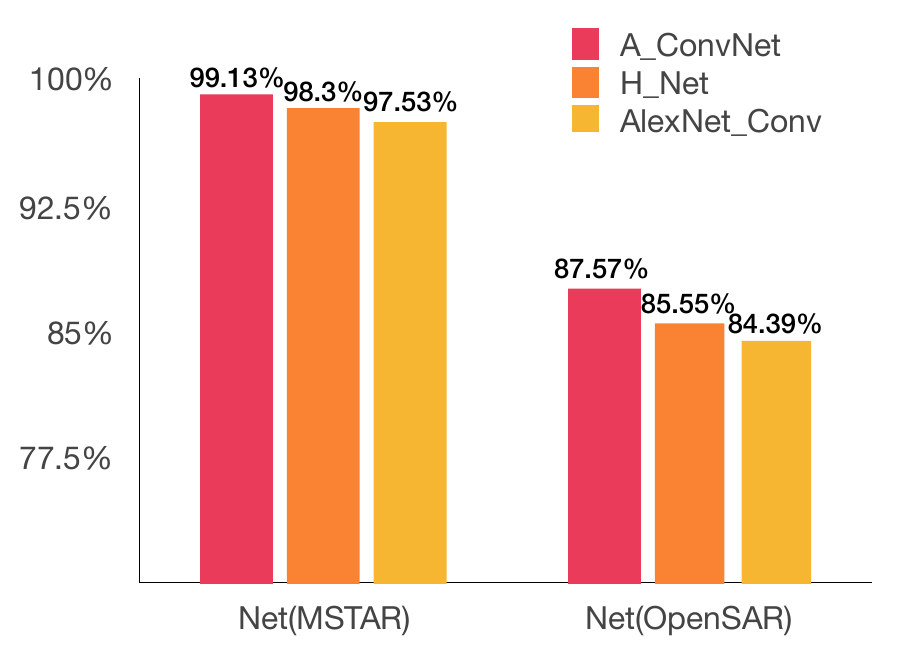}
\caption{Target recognition accuracy on MSTAR and OpenSARShip of A\_ConvNet, H\_Net and AlexNet\_Conv which are all trained from scratch.}
\label{Fig7}
\end{figure}

 As depicted in Table \ref{network}, $a(b)$ denotes $a$ channels and the kernel size of $b\times b$ in each convolution layer. We denote the network $Net$ retrained on MSTAR and OpenSARShip as $Net(M)$ and $Net(O)$, respectively. It can be seen in Fig. \ref{Fig7} that as the network going deeper and wider from A\_ConvNet to AlexNet\_Conv, the performance on SAR targets is decreasing. A\_ConvNet is successful in MSTAR because the smaller network offers an appropriate feature space to fit the limited data. When it comes to a deeper and wider network, a more complex and non-linear function is going to be learnt with limited data which is difficult to find the optimal solution.

\begin{table}[!t]
\caption{The Configuration of Convolution Layers in Three Networks}
\label{network}
\centering
\begin{tabular}{cccc}
\toprule
\textbf{Network} & \textbf{A\_ConvNet} & \textbf{H\_Net} & \textbf{AlexNet\_Conv} \\
\midrule
\textbf{conv1} & 16(5) & 48(5) & 96(11) \\
\textbf{conv2} & 32(5) & 96(5) & 256(5) \\
\textbf{conv3} & 64(5) & 128(3) & 384(3) \\
\textbf{conv4} & 128(6) & 128(3) & 384(3) \\
\textbf{conv5} & None & 256(3) & 256(3) \\
\midrule
\textbf{size} & \textbf{0.4 M} & \textbf{0.7 M} & \textbf{4 M} \\
\bottomrule
\end{tabular}
\end{table}

\begin{figure}[!t]
\centering
\includegraphics[width=8cm]{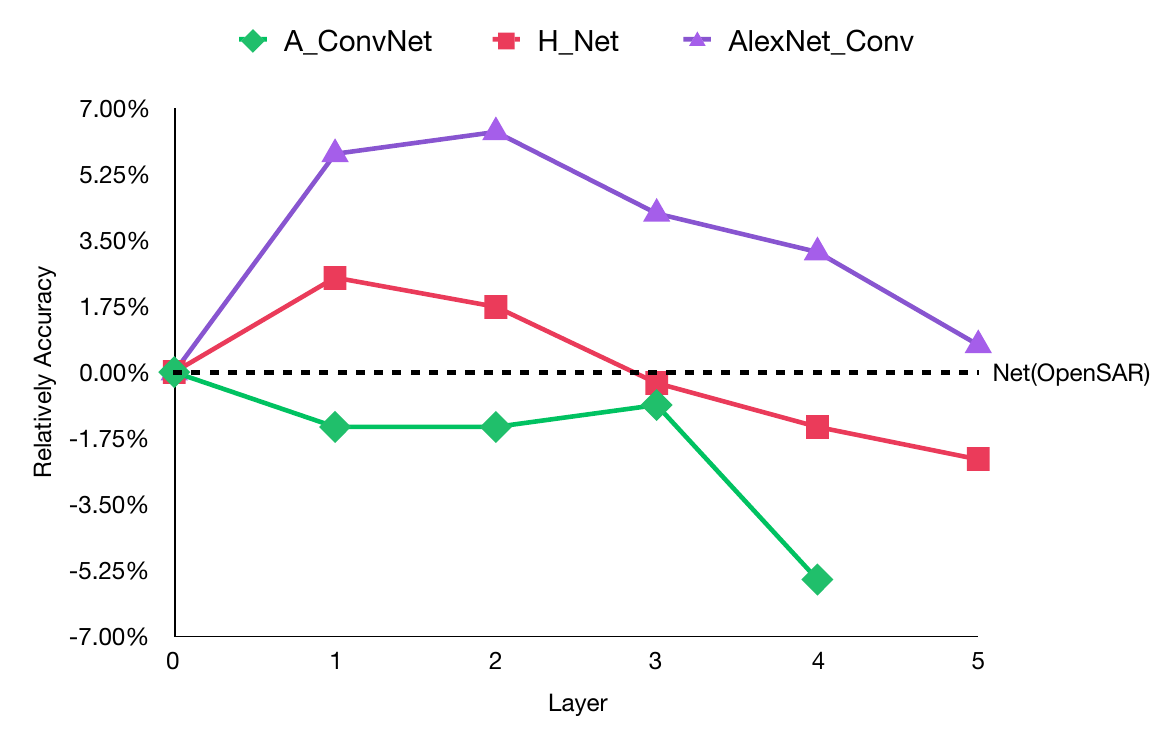}
\caption{The relatively accuracy of $Net(M_{\mathrm{k}}O)$, where $Net$ is from ${\{A\_ConvNet, H\_Net, AlexNet\_Conv \}}$ and $k$ denotes the layer of $Net(\cdot)$}
\label{Fig8}
\end{figure}

Next, we follow the instruction in Section \ref{3.1} to analyze the feature transferability in each layer of the three models, $A\_ConvNet(M)$, $H\_Net(M)$, and $AlexNet\_Conv(M)$. Considering the hyper-parameter of the conv5 layer in A\_ConvNet is specific to classification, only the first four convolution layers are transferred in our experiments. We record the recognition rate on OpenSARShip test data as ${Net(M_{\mathrm{k}}O)}$, denoting the model trained by transferring and freezing the first $\mathrm{k}$ layers of the model $Net(M)$ where $Net$ is in ${\{A\_ConvNet, H\_Net, AlexNet\_Conv \}}$. The remaining higher layers together with the classification layer are randomly initialized and trained on OpenSARShip. The performance of $Net(M_{\mathrm{k}}O)$ is shown in Fig. \ref{Fig8} and Table \ref{transferability}. Although A\_ConvNet performs better on small scale dataset like OpenSARShip and MSTAR training from scratch than H\_Net and AlexNet\_Conv, the features in each layer of $A\_ConvNet(M)$ reflect low generality to OpenSARShip, observing the performance of $A\_ConvNet(M_{\mathrm{k}}O)$ is not as good as $A\_ConvNet(O)$. On the other hand, the over-parameterized networks $H\_Net(M)$ and $AlexNet\_Conv(M)$ improve the performance on OpenSARShip in transfer learning. 

It can be inferred that even though the difficulty for a small dataset to find an optimal solution in training a deeper and wider network, the learnt features are more general to a related task so that the transferring features are able to help the related target task find a better solution.

\renewcommand{\multirowsetup}{\centering} 
\begin{table*}[!t]
	\caption{The OpenSARShip Recognition Rate of Different Networks and Source Tasks When Transferring and Freezing Different Layers}
	\label{transferability}
	\centering
	\begin{tabular}{cccccccc}
		\toprule
		\multirow{2}{*}{\textbf{Network}} & \multirow{2}{*}{\textbf{Source Task}} & \multirow{2}{*}{\textbf{Net(OpenSAR)}} & \multicolumn{5}{c}{\textbf{Transferred Layers (Frozen)}} \\
		\cmidrule{4-8}
									&			& 			 & \textbf{1} & \textbf{2} & \textbf{3} & \textbf{4} & \textbf{5} \\
		\midrule
		\textbf{A\_ConvNet}			& MSTAR 	& 0.8757 	& 0.8612 	& 0.8612 	& 0.8670 	& 0.8208 	& none \\
		\midrule
		\multirow{3}{*}{\textbf{H\_Net}}
									& SAR(recon) 	& \multirow{3}{*}{0.8555} 	& 0.8641 & 0.8483 & 0.8223 & 0.7818 & 0.68	\\
									& MSTAR	&						& 0.8805 & 0.8728 & 0.8526 & 0.8410 & 0.8324 \\
									& SAR(recon)*MSTAR	&					& \textbf{0.8844} & \textbf{0.88}	  & \textbf{0.8858} & \textbf{0.8902} & \textbf{0.8584} \\
		\midrule
		\multirow{3}{*}{\textbf{AlexNet\_Conv}}
									& ImageNet & \multirow{3}{*}{0.8439} & 0.8901 & 0.8584 & 0.8468 & 0.8584 & 0.7774 \\
									& SAR 	&					 	& 0.8974 & 0.8988 & \textbf{0.8883} & \textbf{0.8921} & \textbf{0.8526}	\\
									& MSTAR	&						& \textbf{0.9017} & \textbf{0.9075} & 0.8859 & 0.8757 & 0.8511 \\
									
		\midrule
		\multirow{4}{*}{\tabincell{c}{\textbf{AlexNet\_Conv} \\ (transitive transfer)}}
									& ImageNet*SAR 	& \multirow{4}{*}{0.8439}	& 0.8930 & 0.8901 & 0.8718 & 0.8671 & 0.7109 \\
									& ImageNet*MSTAR	&			& 0.8931 & 0.8902 & 0.8815 & 0.8902 & 0.7283 \\
									& ImageNet*SAR*MSTAR	&		& \textbf{0.8988} & 0.8931 & \textbf{0.8960} & \textbf{0.9017} & 0.7486 \\
									& SAR*MSTAR	&					& \textbf{0.8988} & \textbf{0.9032}	  & \textbf{0.8959} & 0.8872 & \textbf{0.8612} \\
		\bottomrule
	\end{tabular}
\end{table*}

\subsubsection{What Source Data / Tasks}
\label{4.2.2}

Intuitively, we can imagine that the closer data or tasks are better to provide transferable features to SAR target recognition, such as other kind of SAR target recognition, SAR land cover classification. However, in some cases, we do not have enough labeled SAR data to pre-train a deeper network with strongly representative features. On the other hand, the abundant unlabeled SAR images can be easily collected. Huang et al. \cite{huang2017transfer} indicated that the large scale of unlabeled SAR scene data can be reconstructed with training a stacked convolution auto-encoders of which the stacked convolutional layers are capable to transfer to SAR target recognition task. Still, whether the well-known natural images pre-trained models popular in transferring to other remote sensing tasks are transferable to SAR targets remains to be explored.

\begin{figure}[!t]
\centering
\includegraphics[width=9cm]{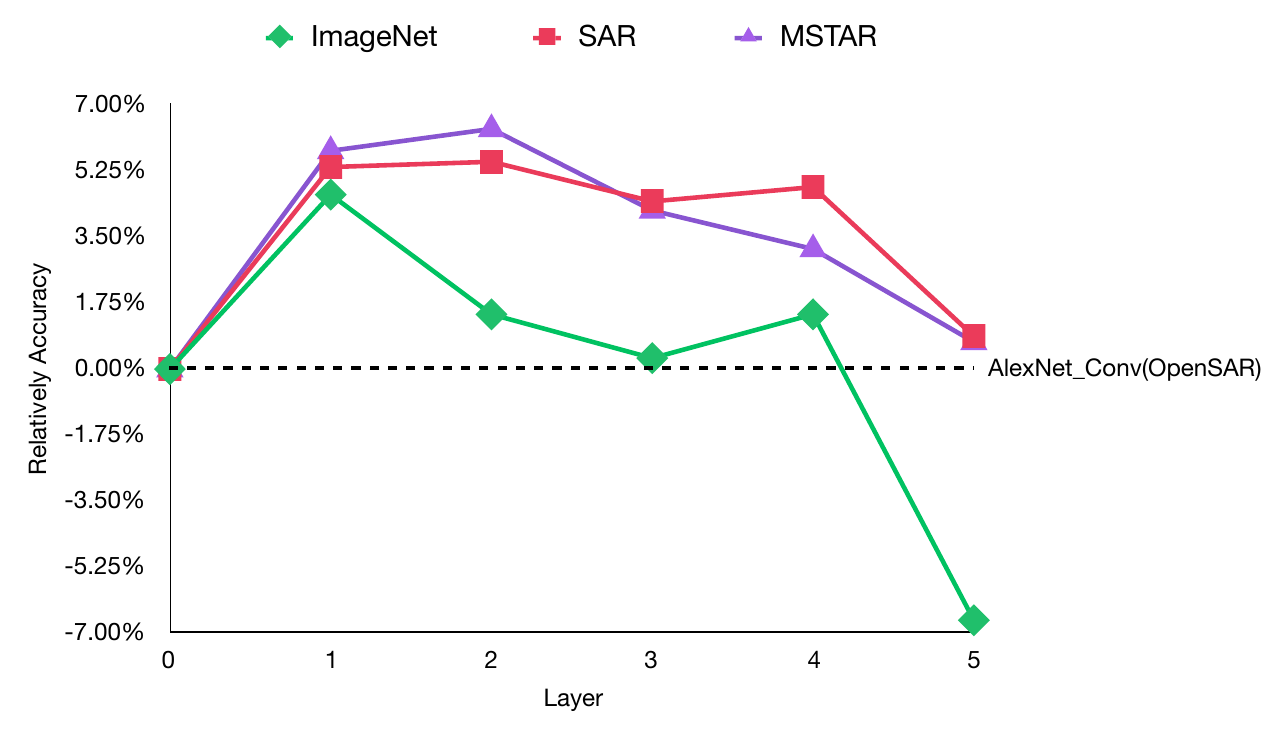}
\caption{The relatively accuracy of $AlexNet\_Conv(I_{\mathrm{k}}O)$, $AlexNet\_Conv(S_{\mathrm{k}}O)$, $AlexNet\_Conv(M_{\mathrm{k}}O)$, respectively, compared with the performance of $AlexNet\_Conv(O)$, where $k$ denotes the layer of $AlexNet\_Conv(\cdot)$. The points above the black baseline indicate the generality of the features in the $kth$ layer and those below the black baseline indicate the specificity.}
\label{what_source1}
\end{figure}

Firstly, we experiment the AlexNet\_Conv pre-trained with ImageNet denoted as $AlexNet\_Conv(I)$, SAR land cover dataset denoted as $AlexNet\_Conv(S)$ and MSTAR denoted as $AlexNet\_Conv(M)$ respectively in transferring to OpenSARShip and the results can be found in Fig. \ref{what_source1} and row 4 of Table \ref{transferability}. Compared with the source tasks of SAR land cover classification and MSTAR target recognition, the features in the first layer of $AlexNet\_Conv(I)$ perform well on generalizing but show much specificity in higher layers, performing a significant drop when transferring and freezing the second to fifth convolution layers. Even though the low-level features learnt from natural images that resemble Gabor filters are effective to represent SAR targets, the features from higher layers are more specific on natural images which indicates more distant the mid-level features of natural images and SAR targets present, much worse in high layers. On the other hand, features in $AlexNet\_Conv(S)$ and $AlexNet\_Conv(M)$ show more robust on generalization to SAR targets. More specifically, the SAR land cover classification trained model performs better in higher layers due to the large scale dataset with abundant SAR image information and the similar task of classification.

What if we don't have the large-scale annotated SAR images to pre-train a deep network? Row 3 of Table \ref{transferability} shows how the unlabeled SAR images performs in transferring. Due to the distance between unlabeled SAR images reconstruction task and the OpenSARship recognition, the transferability of features in $H\_Net(S)$ decreases to be specific just in layer 2 while the MSTAR recognition task much more similar with our target task results in more general features in $H\_Net(M)$.

Limited SAR annotated data in reality, it is not easy to find a source task which is both similar to SAR targets and with a large amount of related data. With features specific to natural images in higher layers of ImageNet pre-trained models and specific to reconstruction tasks of unlabeled SAR pre-trained models, we are going to explore the transitive transfer method with multi-source tasks related to SAR targets to enhance the generality of features in deep networks.

\begin{figure}[!t]
\centering
\includegraphics[width=8cm]{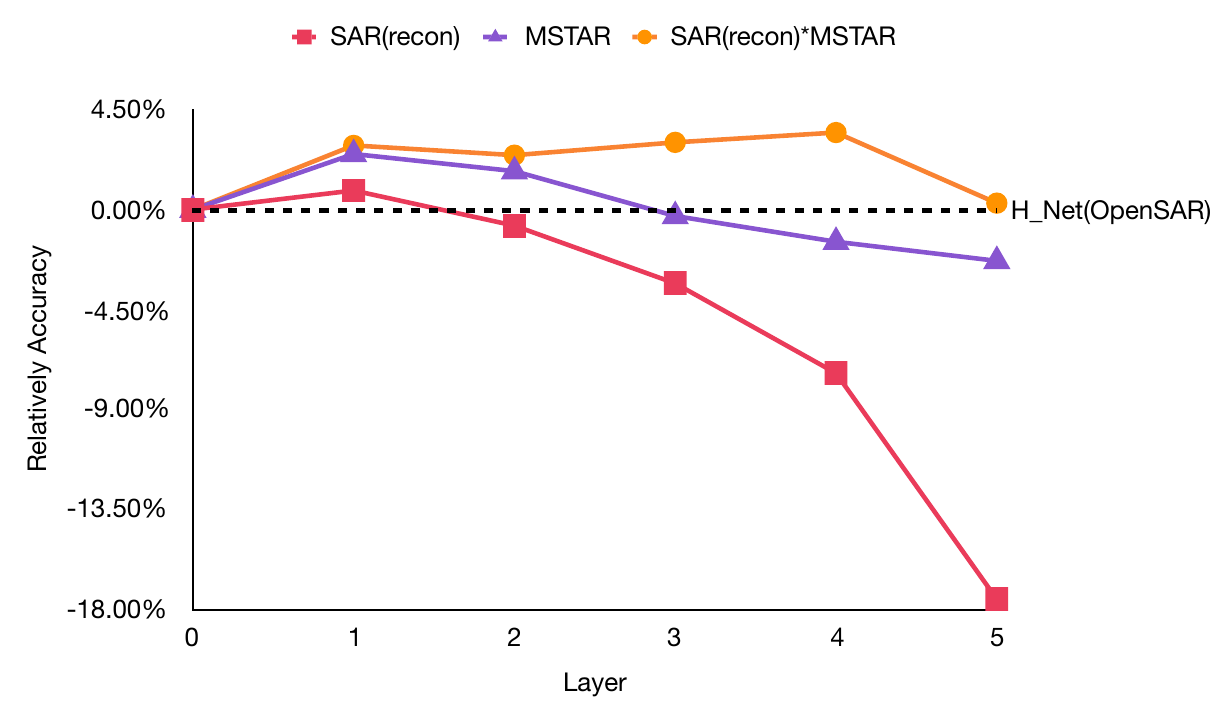}
\caption{The relatively accuracy of $H\_Net(S_{\mathrm{k}}O)$, $H\_Net(M_{\mathrm{k}}O)$, $H\_Net(S*M_{\mathrm{k}}O)$, respectively, compared with the performance of $H\_Net(O)$, where $k$ denotes the layer of $H\_Net(\cdot)$. The points above the black baseline indicate the generality of the features in the $kth$ layer and those below the black baseline indicate the specificity.}
\label{Fig9}
\end{figure}

The $H\_Net(S*M)$ denotes the network of simply fine\-tuning the convolution layers on $H\_Net(S)$ with MSTAR dataset. Fig. \ref{Fig9} shows the performance of transferring different layers of the pre-trained network to OpenSARShip recognition, comparing with the black line which denotes the performance of $H\_Net(O)$. The areas above the black line indicate the features are general to the target task and those below the line indicate the specificity. Strikingly, $H\_Net(S*M)$ distinctly increases the generality of features in mid and high layers which indicates although the distant source task of unlabeled SAR images reconstruction, the intermediate task of MSTAR classification has an impact on enhancing the transferability of features to other SAR target recognition tasks, on the condition that the pre-trained model on SAR images reconstruction provides a good basis.

\begin{figure}[!t]
\centering
\includegraphics[width=8cm]{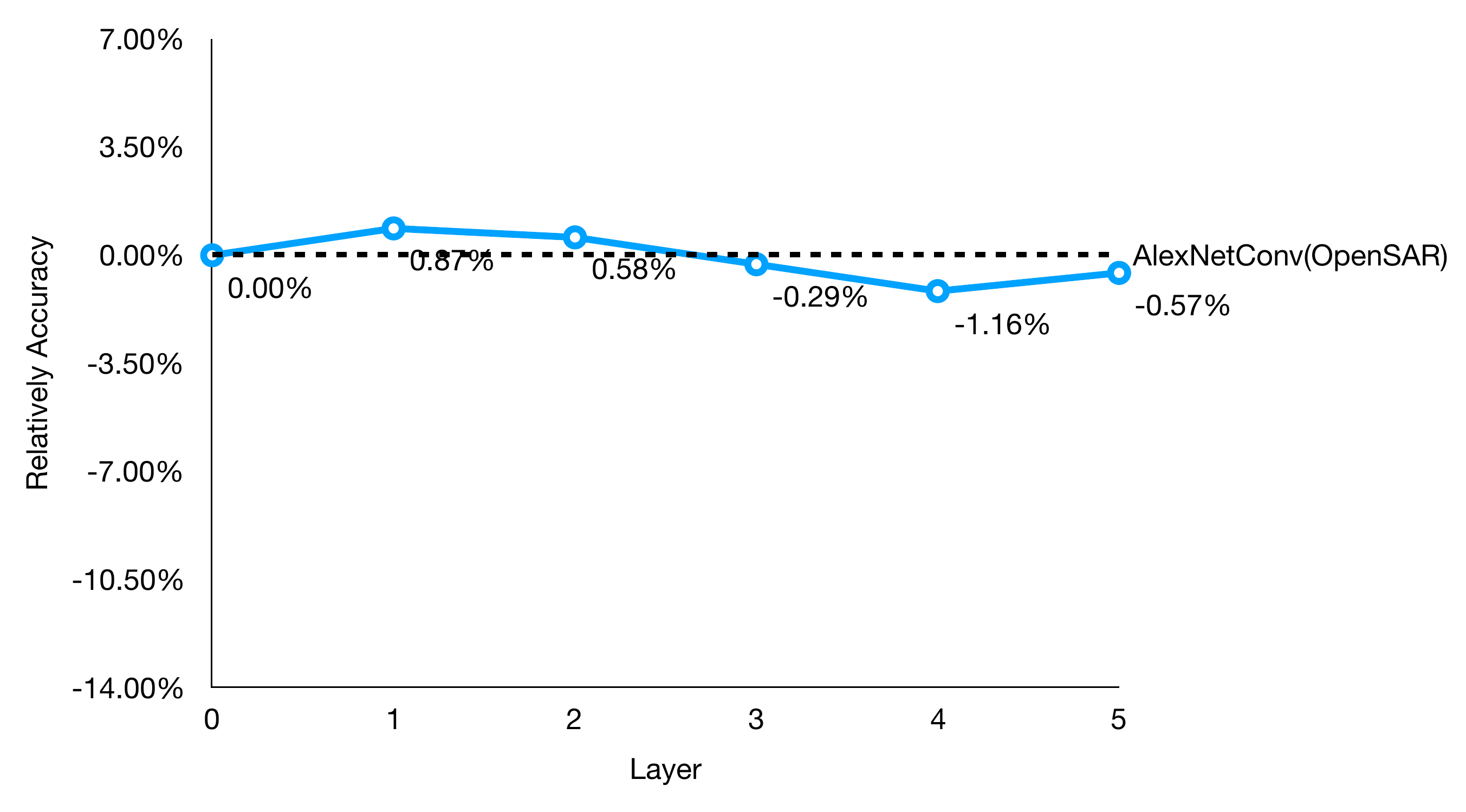}
\caption{The tiny fluctuation of freezing the first $k$ layers of $AlexNet\_Conv(O)$ and training the remaining layers randomly initialized with OpenSARShip dataset. We will ignore the minor influence of co-adaptation in our analysis.}
\label{Fig13}
\end{figure}

\begin{figure}[!t]
\centering
\includegraphics[width=9.5cm]{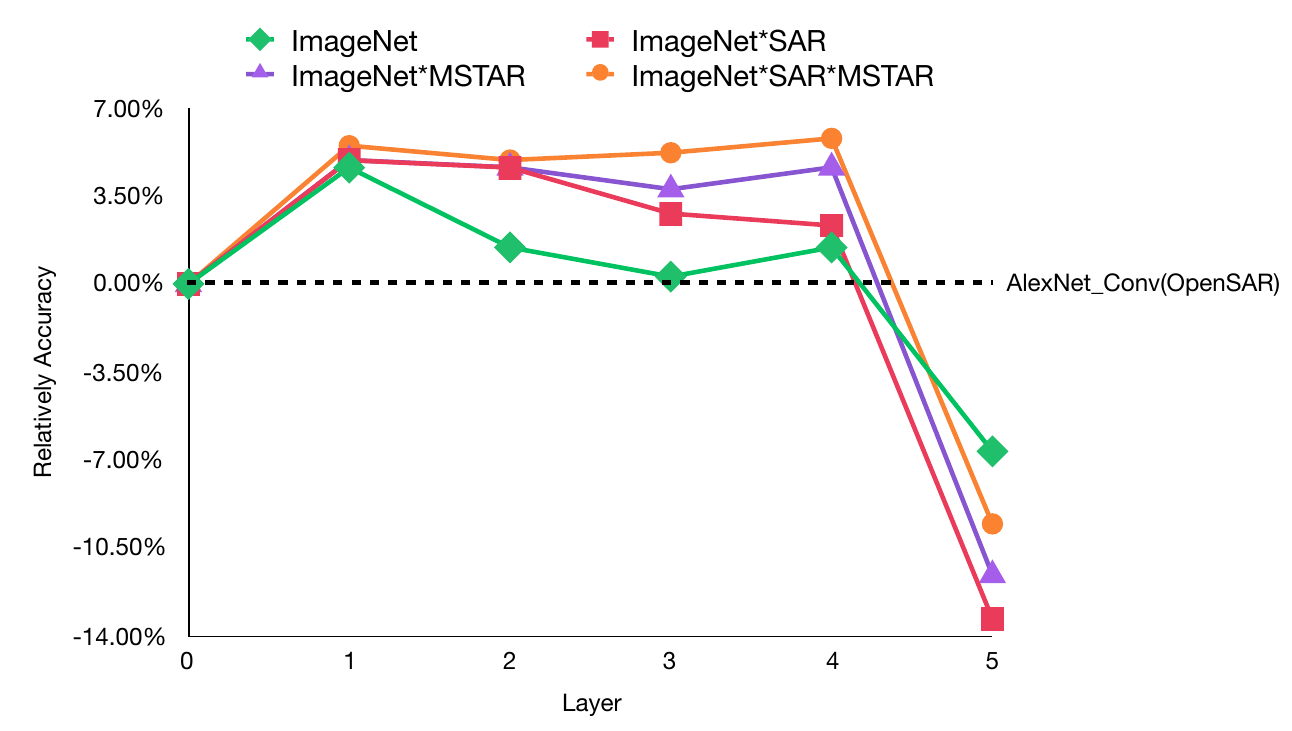}
\caption{Performances on different source tasks of AlexNet\_Conv. The relatively accuracy of $AlexNet\_Conv(I_{\mathrm{k}}O)$, $AlexNet\_Conv(I*S_{\mathrm{k}}O)$, $AlexNet\_Conv(I*M_{\mathrm{k}}O)$, $AlexNet\_Conv(I*S*M_{\mathrm{k}}O)$, respectively, compared with the performance of $AlexNet\_Conv(O)$, where $k$ denotes the layer of $AlexNet\_Conv(\cdot)$. The points above the black baseline indicate the generality of the features in the $kth$ layer and those below the black baseline indicate the specificity.}
\label{Fig10}
\end{figure}

Now that the multi-source transitive transferring performs well on feature generalization, we attempt to explore the ImageNet pre-trained model transferring to SAR target recognition. Yosinski et al. \cite{28yosinski2014transferable} pointed out that the fragile co-adaptation would affect the performances when freezing the first several layers. Our experiments prove that the effect of the co-adaptation in training AlexNet\_Conv with OpenSARShip can be ignored due to the tiny fluctuation, as shown in Fig. \ref{Fig13}. $AlexNet\_Conv(I)$ is fine-tuned with a subset of annotated SAR land cover dataset with 12,000 slices for classification, obtaining $AlexNet\_Conv(I*S)$. Similarly, $AlexNet\_Conv(I*M)$ and $AlexNet\_Conv(I*S*M)$ are obtained with MSTAR dataset. As shown in Fig. \ref{Fig10}, the transferability of features from layer 2 to layer 4 are remarkably increased from $AlexNet\_Conv(I*S)$ to $AlexNet\_Conv(I*M)$, especially in $AlexNet\_Conv(I*S*M)$ where the features generality of the fourth layer is comparable to the lower-level features.

\begin{figure}[!t]
\centering
\includegraphics[width=8cm]{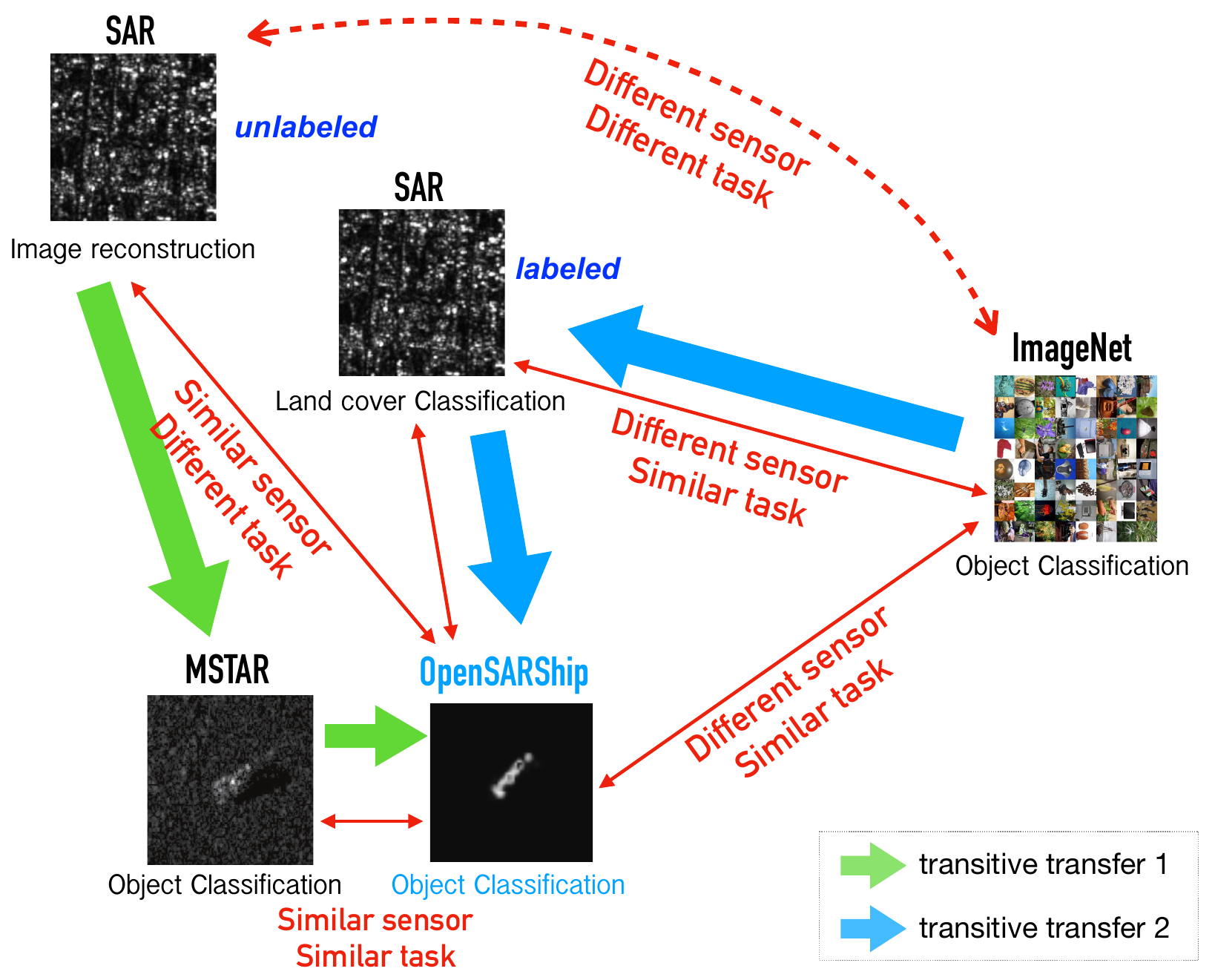}
\caption{The relationship between the target task (OpenSARShip recognition) and different source tasks (ImageNet classification, SAR scene image reconstruction, SAR scene image classification, MSTAR target recognition).}
\label{Fig0}
\end{figure}

The distance among those tasks is illustrated in Fig. \ref{Fig0}. The abundant SAR scene images from similar sensors to SAR targets are suitable for pre-training a deep network to transfer to other SAR related tasks with limited data, but the image reconstruction task with unlabeled data is distant to SAR target recognition task which affect the transferability of features in mid and high layers. As a comparison, the MSTAR classification is close to OpenSARShip recognition while the MSTAR data is limited to train a deeper and wider network. Consequently, if it is possible to obtain a large-scale annotated SAR image dataset, the pre-trained model will be very useful for SAR target recognition. If not, the unlabeled SAR images also help as transitive transfer 1 shows in Fig. \ref{Fig0} where MSTAR classification task can build a bridge between unlabeled SAR image reconstruction and OpenSARShip recognition to improve the generality of features in layers. On the other way, if you want to take the use of the natural images pre-trained models to SAR related problems, we will give an advice to learn some information from SAR images based on the model as transitive transfer 2 shows in Fig. \ref{Fig0}, which is useful in decreasing the specificity to natural images of features in higher layers.

The analysis in this section reveals that the transferability of features is influenced by the generality of the transferred network and the distant between the source and the target tasks, that is to say, the network and the source tasks both have an impact on transferring to SAR target recognition task. Multi-source transitive transferring is a good idea to combine different source datasets from large-scale to limited, as well as from distant to similar, to obtain more general features. The network gradually learns more useful knowledge in the process of completing different tasks. Despite the large diversity between natural images and SAR targets, the low-level features are general and transferable, and fixing with more knowledge of SAR images via multi-source transitive transferring can notably increase the transferability. Multi-source transitive transferring method can not only adopt a larger network, but also combine the greatly generic low-level features of training on ImageNet and the improving transferable features in higher layers.

\subsection{Where and How to Transfer Effectively}

In the previous researches, Yosinski et al. \cite{28yosinski2014transferable} found that the performance drops in fully-connected layers, due to the representation specificity when transferring to other natural images. As a result, the follow-up studies \cite{DANlong2015learning,39oquab2014learning,35tzeng2014deep} are accustomed to adapting features in each of the fully-connected layers when transferring to other natural images. Moreover, Hu et al. \cite{21rs71114680}, Zhao et al. \cite{20zhao2017transfer}, Marmanis et al. \cite{19marmanis2016deep} individually transfer the high-level features from the first fully-connected layer of AlexNet to remote sensing images classification task. However, in our previous discussion we find that this conclusion cannot be simply applied to SAR target recognition. In this section, we will discuss where to transfer features in different situations and how to transfer more effectively to reduce the discrepancy between source and target domain.

\begin{figure}[!t]
\centering
\includegraphics[width=8cm]{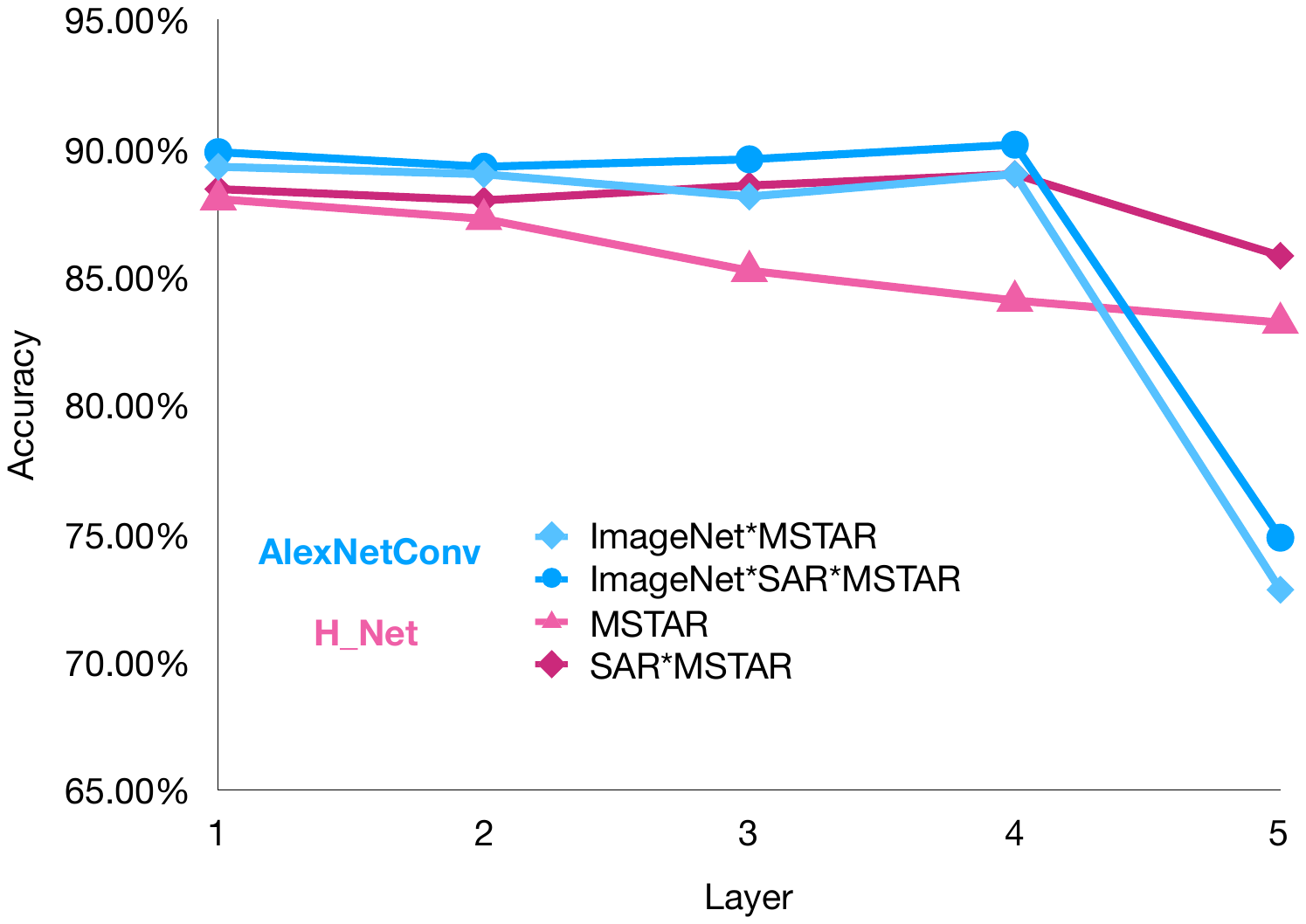}
\caption{The performance of using different source tasks to transfer to OpenSARShip recognition task. The red series lines denote the H\_Net and the blue series lines denote the AlexNet\_Conv.}
\label{Fig12}
\end{figure}

\begin{figure}[!t]
\centering
\includegraphics[width=8cm]{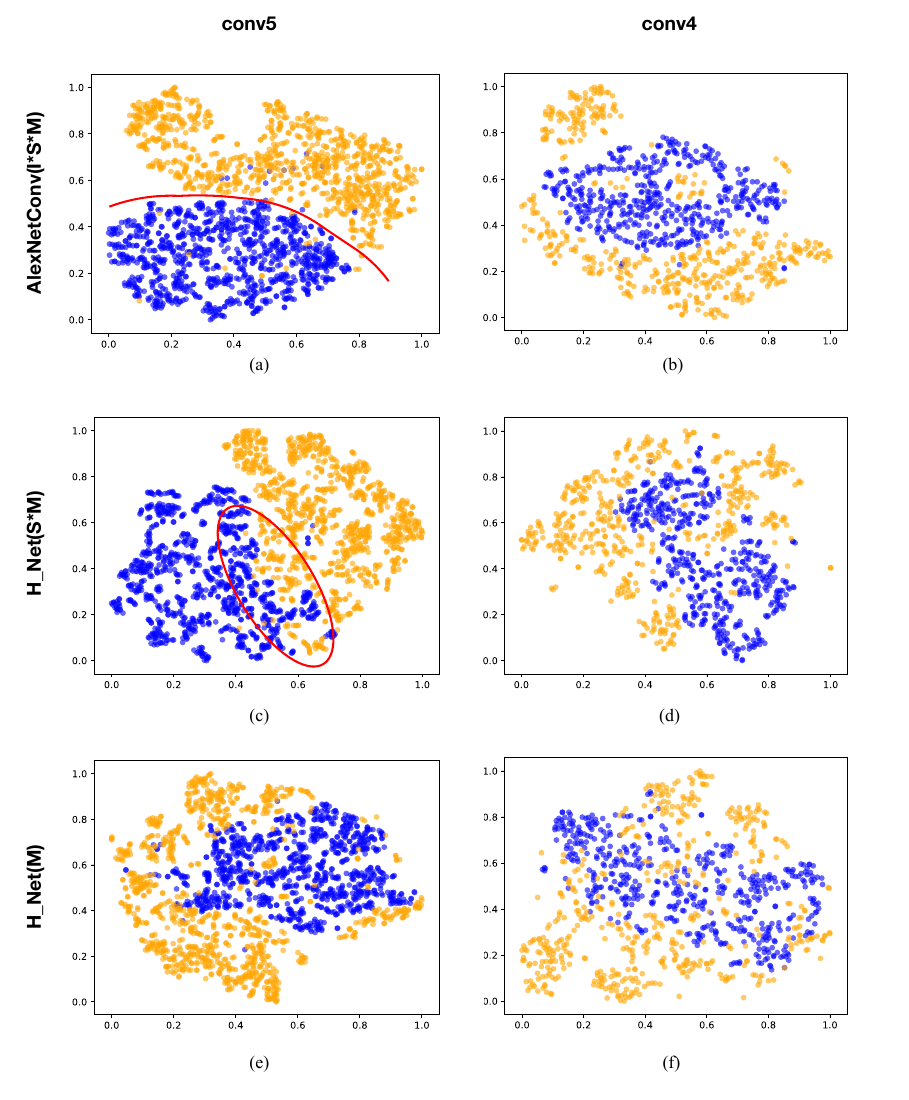}
\caption{The feature visualization of MSTAR and OpenSARShip dataset in layer 5 and layer 4 in different scenarios by t-sne. The blue dots denote the MSTAR and the orange dots denote the OpenSARShip. The (a)(b), (c)(d), (e)(f) represent the network transferred from $AlexNet\_Conv(I*S*M)$, $H\_Net(S*M)$ and $H\_Net(M)$, respectively. The (a)(c)(e) and (b)(d)(f) represent the features from layer 5 and layer 4, respectively.}
\label{Fig15}
\end{figure}

The features in $AlexNet\_Conv(S)$ are good enough to transfer to the SAR targets. For MSTAR dataset, the model achieves an overall accuracy of 99.34\% by fine-tuning all layers, better than the state-of-the-art. And for OpenSARShip, it performs with a fine-tuning result of 91.04\% which is 1.73\%, 1.1\%, 4.23\% and 2.6\% better than $AlexNet\_Conv(I*M)$, $AlexNet\_Conv(I*S*M)$, $H\_Net(M)$ and $H\_Net(S*M)$, respectively. In this part, we mainly focus on the four pre-trained models from AlexNet\_Conv and H\_Net to see how to make them more effective in transferring to SAR targets. Fig. \ref{Fig12} presents the transferability of each layer in different scenarios. In $AlexNet\_Conv(I*S*M)$ and $AlexNet\_Conv(I*M)$ scenario, the first four convolution layers show a strong ability to extract the generic features of OpenSAR but rapidly decreased in layer 5. However, in $H\_Net(M)$ and $H\_Net(S*M)$ scenario, even though the performance of transferring the first four layers are not as good as AlexNet\_Conv ones, the features in layer 5 present a better generalization. We visualize the features in layer 4 and layer 5 from MSTAR and OpenSARShip dataset of different scenarios by t-sne \cite{maaten2008visualizing}, as shown in Fig. \ref{Fig15}, where the blue dots denote the MSTAR dataset and the orange ones denote the OpenSARShip dataset. The features of MSTAR and OpenSARShip of layer 5 in $AlexNet\_Conv(I*S*M)$ can be simply distinguished which indicates the large difference of feature distributions between source and target data. In $H\_Net(S*M)$, however, the feature distribution presents more indistinguishable between source and target data, more noticeable in $H\_Net(M)$. These properties concern the choice of strategies of how to transfer the features in SAR target recognition.

We experiment the ITL and STL algorithms proposed in Section \ref{3.3} in different scenarios of AlexNet\_Conv and H\_Net. 

\subsubsection{AlexNet\_Conv}

\begin{figure}[!t]
\centering
\includegraphics[width=8cm]{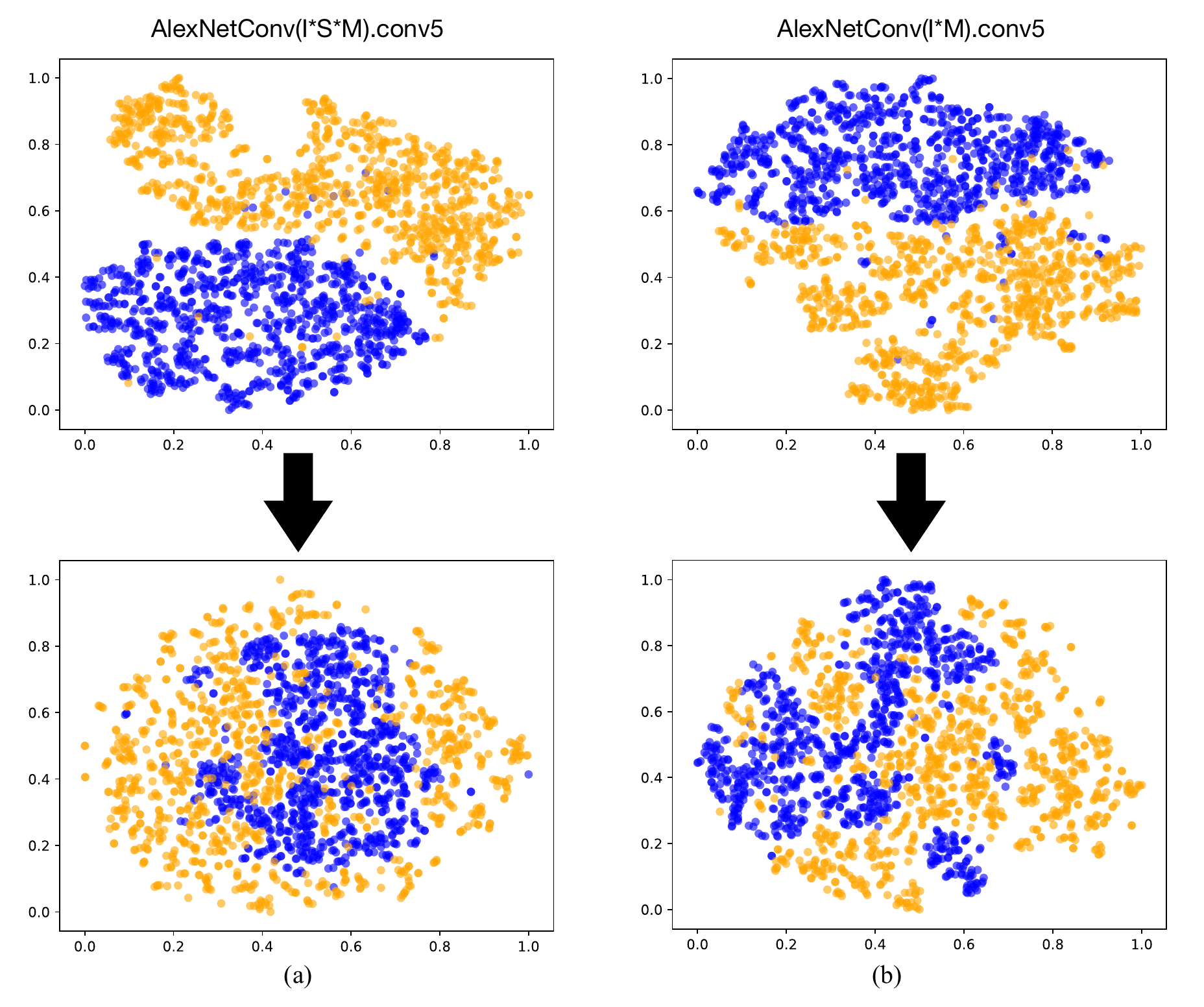}
\caption{The feature distribution in layer 5 becomes more similar after updating the adaptation layers. (a) denotes the $AlexNet\_Conv(I*S*M)$ scenario and (b) denotes the $AlexNet\_Conv(I*M)$ scenario.}
\label{Fig14}
\end{figure}

For STL algorithm, according to the previous analysis, we consider the layer $1\sim4$ as the off-the-shelf layers in $AlexNet\_Conv(I*M)$ and $AlexNet\_Conv(I*S*M)$ because of the great performance of $AlexNet\_Conv(I*M_{\mathrm{4}}O)$ and $AlexNet\_Conv(I*S*M_{\mathrm{4}}O)$ while the layer 5 as the adaptation layer. After the first step of updating the adaptation layer, we can observe an obvious improvement on similar feature distributions of source and target data which implies the discrepancy of features in source and target is decreased, as shown in Fig. \ref{Fig14}. In this step, the off-the-shelf layers are fixed since the quality of generality makes them possible to extract the off-the-shelf features of OpenSARShip data. Moreover, it is an unsupervised learning part so that all labeled and unlabeled OpenSARShip data can be used to narrow the gap of feature distributions. Next, the classification layer is trained with labeled OpenSAR data by combining the cross-entropy loss of labels and outputs of Softmax layer and the transfer loss. In this part, the learning rate in layer 1, 2, 3, 4, 5 are set to $10^{-4}$ so that the previous layers are slightly fine-tuned and the learning rate of classification layer is set to $10^{-2}$ which is 100 times larger than previous layers. The transfer loss constrains the whole network to maintain the property of narrowing the discrepancy between source and target, and it should be controlled by the trade-off $\lambda$ to avoid dominating the total loss and preventing the continuous decreasing of classification loss. In our experiments, we set $\lambda$ as 1.5 in AlexNet\_Conv scenarios. 

Table \ref{how} shows the performance of different algorithms in different scenarios. STL approach boosts the performance by 1.44\% and 1.96\% respectively on $AlexNet\_Conv(I*M)$ and $AlexNet\_Conv(I*S*M)$ compared with the common fine-tuning methods on transfer learning. For ITL approach of combining the transfer loss and classification loss to fine-tune all layers, the results are not as good as the STL but still better than simply fine-tuning method, improving 0.57\% and 0.48\% on $AlexNet\_Conv(I*M)$ and $AlexNet\_Conv(I*S*M)$, respectively. It refers that the transfer loss certainly has an impact on improving the performance of classification but the first 4 layers in AlexNet\_Conv scenarios are enough to extract the general features of OpenSARShip so that the constraint of transfer loss would be better not to affect the off-the-shelf layers to restrict the good feature representation. This also verifies the advantage of analyzing the transferability of layers to distinguish the off-the-shelf and the adaptation layers.

\subsubsection{H\_Net}

In H\_Net scenario, the features of OpenSARShip in layer 4 and layer 5 are more likely to share the similar distribution with MSTAR than AlexNet\_Conv scenarios but the performance of either transferring and fixing the first 4 layers or fine-tuning all layers is worse, as shown in Table \ref{transferability} and \ref{how}. The underlying reason lies in the fact that the base network of $AlexNet(I)$ has a stronger ability to extract generic features than $H\_Net(S)$ and $H\_Net(M)$. Consequently, in lower layers the transitive transferring via multi-source makes an effort of generalizing the features to OpenSAR to improve the performance of SAR target recognition task while in high-level layer the discrepancy dominates the transferability. In our experiments, the ITL approach improves the performance by 1.15\% compared with fine-tuning all layers. With STL algorithm, considering the decline of feature generalization in higher layers, we set the layer 4 and layer 5 as the adaptation layers in scenario $H\_Net(M)$ and improve the performance of 2.02\%, compared with simply fine-tuning all layers. The multiples of learning rate in each convolution layer are set to 0.1, 0.1, 0.1, 0.5, 1, and 10 in classification layer. The results are sensitive to the trade-off value of $\lambda$. In the second step of combining the transfer loss and classification loss, $\lambda$ must be set to a smaller value to constrain the transfer loss due to the major effect on fine-tuning with the classification loss. 

When it comes to $H\_Net(S*M)$, we observe in Table. \ref{how} that the performance of STL is not as good as ITL. Fig. \ref{Fig12} shows that the generalization of features in lower layers is not as good as the layer 5 especially in bottom layers which indicates that the lower layers have the potential to improve the ability of extracting good features by fine-tuning rather than treated as the off-the-shelf layers. ITL and STL improve the performance of recognizing the OpenSARShip by 1.44\% and 0.87\% in $H\_Net(S*M)$, respectively. As a result, combining the transfer loss with the classification loss to fine-tune all layers as ITL approach is a better choice in $H\_Net(S*M)$.

\begin{table}[!t]
	\caption{The performance of different scenarios by using ITL and STL, compared with fine-tuning all layers.}
	\label{how}
	\centering
	\begin{tabular}{ccccc}
		\toprule
		\textbf{Network}	& \multicolumn{2}{c}{AlexNet\_Conv}	& \multicolumn{2}{c}{H\_Net} \\
		\midrule
		\textbf{Source Tasks}	& $I*M$		& $I*S*M$ 			& $M$		& $S$(recon)$*M$			\\
		\midrule
		\textbf{Fine-tune}	& 89.31\%	& 89.94\%			& 86.41\%	& 88.44\%		 \\
		
		\textbf{ITL}				& 89.88\%	& 90.46\%			& 87.28\%	& \textbf{89.88\%}		\\
		
		\textbf{STL}				& \textbf{90.75\%}	& \textbf{91.9\%}		& \textbf{88.43\%} 		& 89.31\%		\\
		\bottomrule
	\end{tabular}
\end{table}

\section{Conclusion}
\label{5}

In this paper, we elaborately explore what network and source tasks are better to transfer, in which layer the features are more generic to transfer and how to effectively transfer in SAR target recognition. We find that the transferability is up to generalization capacity of the network and the distance between source and target task. A small network is appropriate to train with limited labeled SAR targets but when transferring to other SAR target recognition tasks the feature generality is not enough to extract a good representation. As a result, a larger network trained with a large-scale dataset and a source task similar to SAR target recognition are both required. If possible, a deep network pre-trained with a large-scale annotated SAR scene dataset is a good source to transfer and we have released the resource in \cite{code}. Otherwise, the unlimited unlabeled SAR images are also helpful especially using transitive transfer proposed in this paper to transfer knowledge from large-scale dataset to small-scale one, with closer distance to SAR target recognition task. We do not suggest to use natural images pre-trained model straightforwardly to SAR targets due to the large difference between them which may result in much specific features in higher layers. Instead, the mid level features specific to natural images can be generalized to SAR target by transitive transfer with SAR related tasks. In order to decrease the discrepancy between source and target domain in very high layer, the proposed MK-MMD based transfer method to separately train the adaptation layer and slightly update the off-the-shelf layers is recommended which improves the performance than simply fine-tuning all layers in SAR target recognition transferring.


%



\section*{Acknowledgment}

We thank Dr. Corneliu Octavian Dumitru in German Aerospace Center (DLR) to provide the TerraSAR-X annotated land cover images and we also thank Science Service System for the provision of images (Proposals MTH-1118 and LAN-3156).

\ifCLASSOPTIONcaptionsoff
  \newpage
\fi



%


\bibliographystyle{IEEEtran}
\bibliography{IEEEabrv,preprint}

\end{document}